\documentclass[numberedappendix]{emulateapj}

\newcommand{\thetabold}{\mbox{\boldmath$\theta$}}
\newcommand{\sigmabold}{\mbox{\boldmath$\sigma$}}
\newcommand{\B}{Bayes\textsc{Clumpy}}

\begin{document}

\title{\B: Bayesian Inference with Clumpy Dusty Torus Models}
\author{A. Asensio Ramos \& C. Ramos Almeida}

\affil{Instituto de Astrof\'{\i}sica de Canarias, 38205, La Laguna, Tenerife, Spain}
\email{aasensio@iac.es}

\begin{abstract}
Our aim is to present a fast and general Bayesian inference framework based on the synergy between machine
learning techniques and standard sampling methods and apply it to infer the physical properties
of clumpy dusty torus using infrared photometric high spatial resolution observations of active galactic nuclei.
We make use of the Metropolis-Hastings Markov Chain Monte Carlo algorithm for sampling the
posterior distribution function. Such distribution results from combining all a-priori
knowledge about the parameters of the model and the information introduced by the observations. The main 
difficulty resides in the fact that the model used to explain the observations is computationally
demanding and the sampling is very time consuming. For this reason, we apply a set of
artificial neural networks that are used to approximate and interpolate a database of models.
As a consequence, models not present in the original database can be computed ensuring continuity.
We focus on the application of this solution scheme to the recently developed public
database of clumpy dusty torus models.
The machine learning scheme used in this paper allows us to generate any model from the
database using only a factor $10^{-4}$ of the original size of the database and a
factor $10^{-3}$ in computing time. 
The posterior distribution obtained for each model parameter allows us to investigate how
the observations constrain the parameters and which ones remain partially or completely undetermined,
providing statistically relevant confidence intervals.
As an example, the application to the nuclear region of Centaurus A shows that the optical depth of the clouds,
the total number of clouds and the radial extent of the cloud distribution zone are well constrained using only 
6 filters. The code is freely available from the authors.
\end{abstract}




\section{Introduction}
It is customary that physical information about astrophysical objects cannot be obtained directly
from the observables. In such a case, astrophysicists propose a plausible scenario 
described by a physical model and the procedure is to compare the observables with the
predictions of the model with the aim of inferring the physical parameters of the model.
The presence of degeneracies (either induced by the presence of noise or intrinsic to the
model) introduce complexity in the analysis and they need to be taken into account. This is
the subject of Bayesian data analysis that, although it is rooted on ideas 
developed in the 19th century, it has become practical only in the last decades.

The fundamental idea behind Bayesian data analysis is to take into account that all
parameters of a model can be understood as random variables with associated
probability distribution functions. The standard problem of model fitting is usually
seen as finding the set of model parameters that better reproduce the observables. However,
the Bayesian approach is far more informative and transforms the problem into 
finding the probability distribution
function associated with the parameters of the model once the data set is taken into account. In the presence of noise
and/or degeneracies, these probability distribution functions represent the 
complete solution to the problem and automatically
include all the statistical information about the parameters that can be inferred from
the observables. We have witnessed an enormous interest in Bayesian inference in 
Astrophysics in the last decade. The reason for this resides in two facts. First, the quality and amount
of observed data is usually deficient and one has to rely on methods that
exploit to the limit the reduced amount of information. Second, the 
applied physical models are sometimes too complex as compared with the data available
to constrain them. To mention a few recent works, we
find applications in cosmological analyses 
\citep[e.g.,][]{lewis02,rubino_martin03,rebolo04,trotta08}, gravitational wave analyses 
\citep[e.g.,][]{cornish05}, gravitational lensing \citep[e.g.,][]{brewer_lensing07}, 
oscillation of solar-like stars \citep[e.g.,][]{brewer_oscillations07}, analysis of
spectropolarimetric data \citep{asensio_martinez_rubino07}, analysis of extreme ultraviolet 
spectral line fluxes \citep{kashyap98}, and more.

As we review in Appendix \ref{sec:bayesian_inference}, Bayesian inference techniques can 
be essentially reduced to the calculation of multi-dimensional integrals 
\cite[e.g.,][]{neal93,skilling04,gregory05,trotta08}. In very simple models, these integrals can 
be carried out analytically. However, more realistic problems cannot be analytically
treated. The explosion of Bayesian analysis methods in the last decades has to be
associated with the set of
efficient sampling techniques today known as Markov Chain Monte Carlo methods
\citep[MCMC;][]{metropolis53,neal93,gregory05}. In spite of their success, these methods also present
the drawback of being computationally intensive because the proposed
model has to be evaluated many times. As a consequence, the execution time of these 
techniques is quite high if the evaluation time of the model is non-negligible. For this reason, there have been
some efforts in recent years towards reducing the evaluation time of the models 
at the expense of a small lost in accuracy. They are based on the development of
approximate methods that are able to ``learn'' a database of models for many combinations
of the model parameters. For instance, 
\cite{fendt_pico07} developed a method based on polynomial interpolation for the rapid
cosmological parameter estimation problems. Later, \cite{auld08} used a neural network
approach for the calculation of cosmic microwave background power spectra (only
for models in a small hypercube around the commonly accepted region of most probable
values for the cosmological parameters) leading to
a very fast Bayesian cosmological parameter estimation code.

Our main aim in this paper is to present \B, a computer program that allows us to efficiently
carry out Bayesian analysis of observed spectral energy distributions coming from the
inner region of active galactic nuclei (AGN). To this aim, we use the recently developed clumpy dusty torus model of 
\cite{Nenkova08a,Nenkova08b}, known as CLUMPY models, and develop a MCMC code whose output is the probability
distribution function for all the parameters of the CLUMPY models once the 
observations are taken into account. As a consequence, the code yields statistically significant estimations 
of the parameters and, more important, statistically relevant confidence intervals. 
This facilitates the investigation of degeneracies and can be also used to suggest
future observations that can help us introduce stronger constraints in the inference.
The code is based on a recently released on-line database of CLUMPY 
models\footnote{\texttt{https://newton.pa.uky.edu/$\sim$clumpyweb/}}. We apply an interpolation
method like that presented by \cite{auld08} that greatly accelerates the evaluation
of models. In our case, we manage to make the approximation method work correctly 
for the whole database and not only for a small hypercube, thus allowing us to efficiently 
explore the full space of parameters. Finally, the Bayesian character of the approach allows
the user to include any a-priori knowledge about any parameter.

\section{CLUMPY models}
\label{sec:clumpy_models}
According to the Unified Model for Seyfert galaxies \citep{Antonucci93,Urry95}, 
Type-2 AGN are the edge-on counterparts of the face-on Type-1 AGN. This way, 
in Type-1 AGN the broad-line region (BLR), that is surrounded 
by a dusty torus of a few parsecs \citep{Tristram07}, is observed directly, together with the 
narrow-line region (NLR) emission, whereas in the case of 
Type-2 AGN, only the NLR emission is seen directly. 
However, the Unification Model is not universally 
applicable, since there are several galaxies that do not reveal the broad lines in polarized 
light. 

Regardless, it is clear that there is dust surrounding the central region 
of AGN distributed in a toroidal shape. The dust grains in the torus absorb the 
ultraviolet photons from the 
central engine and, after reprocessing the radiation, are re-emitted in the 
infrared range.
Since many of the predictions of the first compact-torus models 
\citep{Pier92,Granato94,Efstathiou95,Granato97} have not been confirmed by the observations, 
the search for a more distributed or complex geometry of the absorbing material 
around the AGN have been promoted \citep{Nenkova02,Fritz06,Elitzur06,Ballantyne06}.

The clumpy dusty torus models \citep{Nenkova02,Nenkova08a,Nenkova08b,Honig06,Schartmann08} propose that the dust 
is distributed in clumps, instead of homogeneously filling the torus volume. As an example of the
success of these models, they permit to explain, for example, the observed mutations between Type-1 and Type-2 objects detected in the
spectra of a few AGN \citep{Aretxaga99,trippe08}.

Since the reprocessed radiation from the torus is emitted in the infrared, this range is key 
to put constrains to the clumpy dusty torus models. High-resolution observations at these
wavelengths are mandatory, due to the small size of the torus \citep{Tristram07}. This way, 
it is possible to separate the nuclear emission from that of the surrounding galaxy.
Important observational constraints for the torus models come then from the shape of the infrared 
spectral energy distributions (SEDs). Accuracy in the photometry, a filter set spanning a broad 
wavelength range, and well-sampled SEDs are required to restrict the model parameters.

The CLUMPY models that we use in this work \citep[described in][]{Nenkova08a,Nenkova08b} consist of a clumpy distribution of 
matter with a radial extent characterized by $Y = R_\mathrm{o}/R_\mathrm{d}$, that is the ratio of the outer to the inner
radii of the toroidal distribution. The inner radius ($R_\mathrm{d}$) is defined by the dust sublimation
temperature. 
Each clump is specified by its optical depth ($\tau_{V}$), and all clumps are assumed to have the same
optical depth. The dust extinction profile corresponds to a standard cold/oxygen-rich ISM dust
\citep{Ossenkopf92}.
These clumps, of a given dust composition, are heated by an AGN with a given spectral shape and
luminosity. Thus, the inner radius ($R_\mathrm{d}$) is determined uniquely by the luminosity and the chosen dust
sublimation temperature.
The number of clouds along a radial equatorial path is defined as $N$. The radial density profile is a
power law ($r^{-q}$), with an angular distribution characterized by a width parameter, $\sigma$.

\begin{figure*}
\plotone{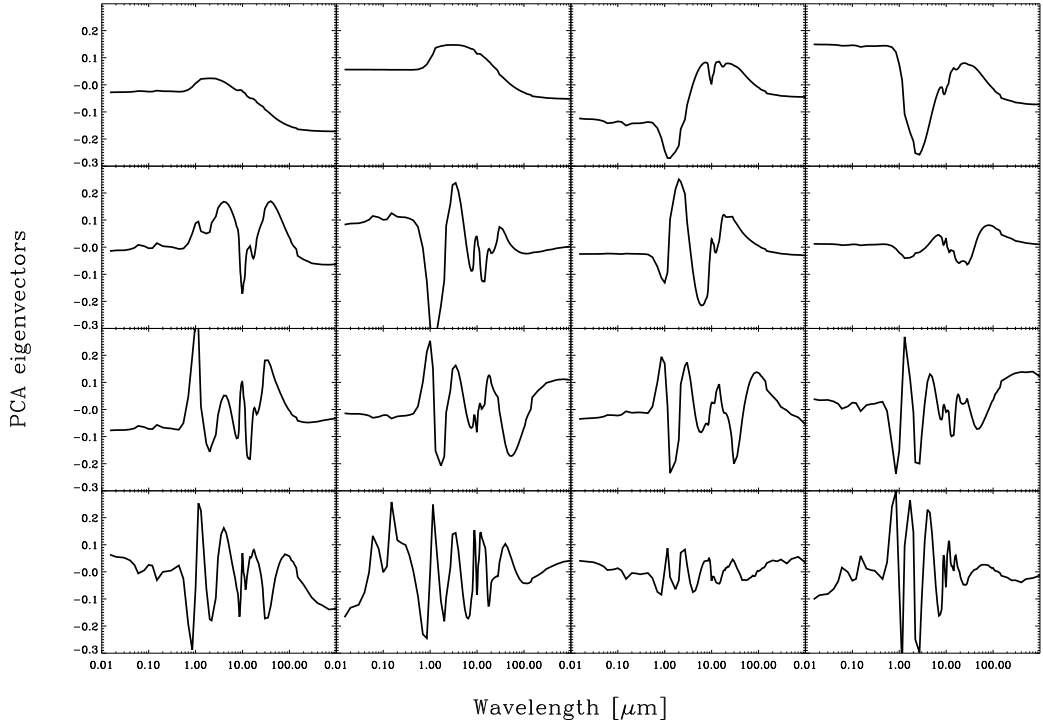}
\caption{First 16 PCA eigenvectors obtained from the CLUMPY database. We have demonstrated that 
the full CLUMPY database can be decomposed using only the first 13 eigenvectors with great
precision. The wavelength variation of the standard deviation of the difference between the 
original models and the truncated reconstruction is shown in Fig. \ref{fig:error}.}
\label{fig:PCA_evecs}
\end{figure*}

\section{\B}
Every CLUMPY model requires several seconds to be calculated. For the typical
lengths of converged Markov Chains, this would amount to something between 1 and 
6 days per Markov Chain for only one run of the inference problem. Obviously, this
is something that is absolutely unacceptable if one wants to carry out inference
for many AGN.

The computational efforts carried out by the CLUMPY group has allowed them 
to calculate and distribute for 
public access around 2$\times$10$^5$ models (now increasing to more than 10$^6$ models).
They have been calculated for a quite fine grid of model parameters.
This database can be used to overcome the difficulty of evaluating the CLUMPY models by
using it for interpolation. The reason for interpolation is that the 
MCMC code will propose models that are not present in the original grid, so that an 
efficient interpolation method has to be applied. If the interpolation method is fast
enough, it will allows us to carry out systematic studies of the compatibility of model 
parameters with different observations in a completely Bayesian framework. Studies like 
analyzing the amount of information added by a given filter and which parameters can be confidently 
recovered from the data are possible under this framework.

In the following sections we describe our approach for interpolation. It is based on the 
application of two different machine learning techniques: principal component analysis (PCA)
for dimensionality reduction and artificial neural networks (ANN) for the
interpolation. Such a method has been already applied by \cite{auld08} for approximate Bayesian inference of
cosmological parameter in a small hypercube around the commonly accepted values of the
cosmological parameters. A similar approach has also been employed by \cite{carroll_fast08} for the fast 
synthesis of Stokes profiles in 
magnetic atmospheres and the quick solution of Zeeman-Doppler imaging problems.

\subsection{Principal Component Analysis}
Each SED in the database is sampled at $N_\lambda=124$ wavelength points. Clearly,
some correlations exist between different wavelength points, so that when the flux
at a given wavelength is modified, the surrounding wavelength points are also 
modified in a very similar way (continuity of the SED). As a consequence, the dimension 
of the non-linear manifold in which the SEDs ``live'' is much smaller than 124 \cite[see][]{asensio_dimension07}.
This fact can be harnessed to apply dimensionality reduction techniques and efficiently compress
the database. Although many complex technique exist, we apply here a very basic linear dimensionality 
reduction technique based on the Principal Component Analysis \cite[PCA; see][]{loeve55} also known as 
Karhunen-Lo\`eve transformation. Briefly, the idea is to obtain a self-consistent basis (principal
components) in which the data can be efficiently developed. This basis has the property that the 
largest amount of variance is explained with the least number of basis functions. It is 
useful to reduce the dimensionality of data sets because most part of the variability of the signal is
carried by the first $N' \ll N_\lambda$ eigenvectors. Note that, since PCA performs a
linear analysis, it is not possible to reduce the dimensionality of the transformed
manifold to the real dimensionality of the non-linear manifold and the number of necessary eigenvectors
is larger than the number of physical parameters of the model \citep{asensio_dimension07}. See
appendix \ref{sec:appendix_pca} for more technical details on PCA.

\begin{figure}
\centering
\plotone{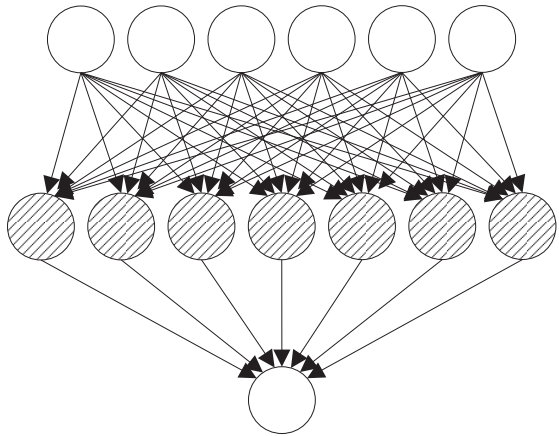}
\caption{Topology of the artificial neural networks applied in this work. The neural networks contain 
an input layer of six neurons for the six fundamental parameters of the CLUMPY models. The output
layer is composed of only one neuron that is associated with the projection of the SED along
each PCA eigenvector. The intermediate layer (widely known as ``hidden'' layer) is used to
obtain the non-linear mapping between the input and the output layers.}
\label{fig:neural_network}
\end{figure}

The first 16 PCA eigenvectors obtained from the
CLUMPY database are shown in Fig. \ref{fig:PCA_evecs}. This figure shows that 
low-order eigenvectors are very smooth and take into account large-scale variations
that are seen in the majority of the SEDs. On the contrary, high-order eigenvectors 
contain high-frequency details that produce small-scale details in a small
amount of SEDs. Choosing only the first $N'=13$ eigenvectors results in a very
good representation of the whole database. In other words, this allows us to reduce the
size of the database because we only need to give 13 numbers for each SED (the projection of
each SED along each PCA eigenvector) and their associated eigenvectors. Note that, for very large
number of models $>10^5$, the reduction in size of the database tends to $13/N_\lambda$, which
is close to 1/10 in our case.

Although the PCA eigenvectors can be usually associated to different physical mechanisms
\citep[e.g.,][]{skumanich02}, our aim here is not to analyze them. We treat the PCA eigenvectors
as a basis set of purely mathematical character that allows us to efficiently develop the
database. Efficiency in our case means that the number of eigenvectors needed to reproduce
the database with a given error bar is the smallest possible. In any case, it is possible to 
see some well-known signatures like the 10 \micron\ one produced by dust emission/absorption in
some eigenvectors of Fig. \ref{fig:PCA_evecs}.

\begin{figure*}[!t]
\plottwo{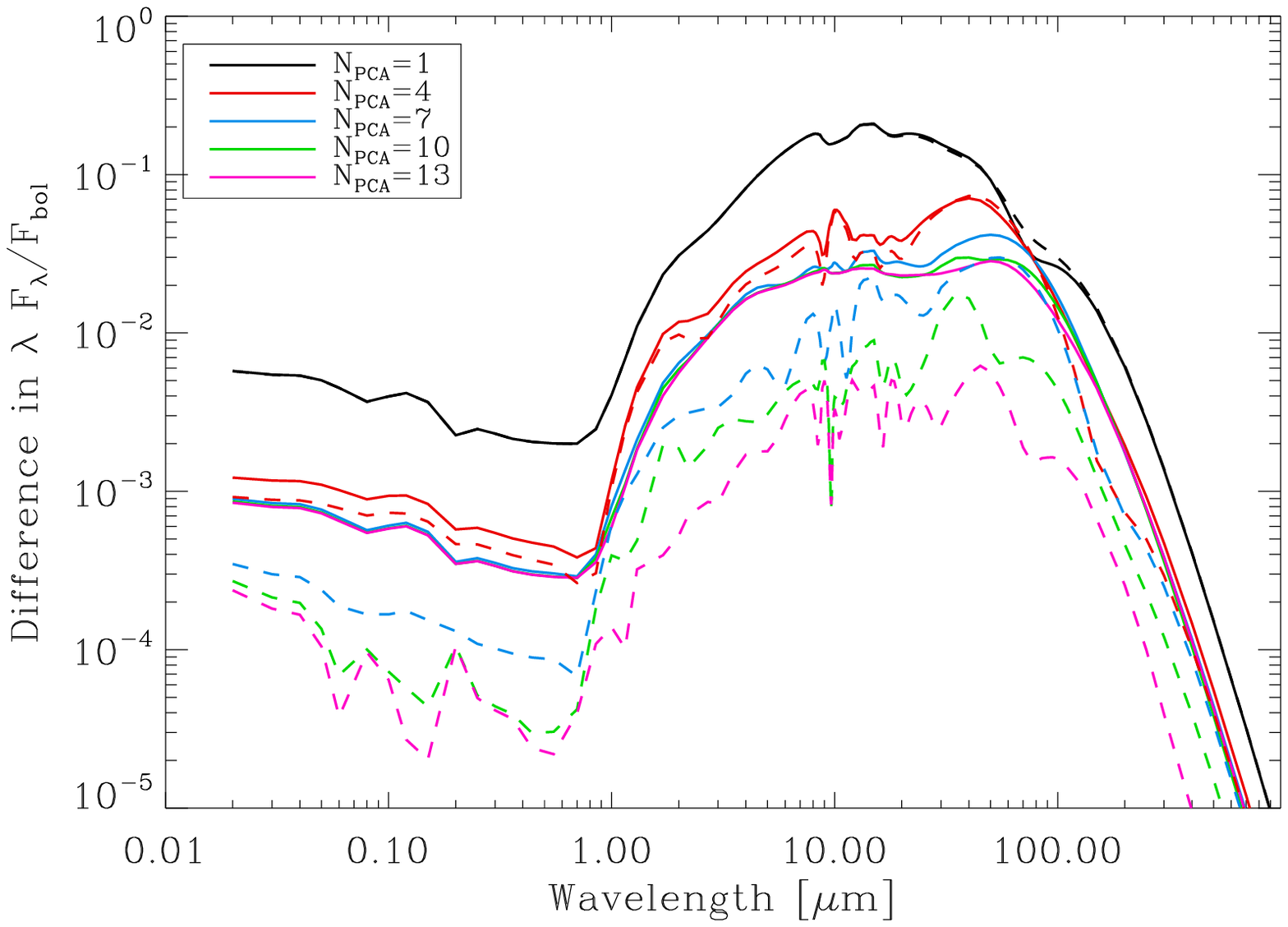}{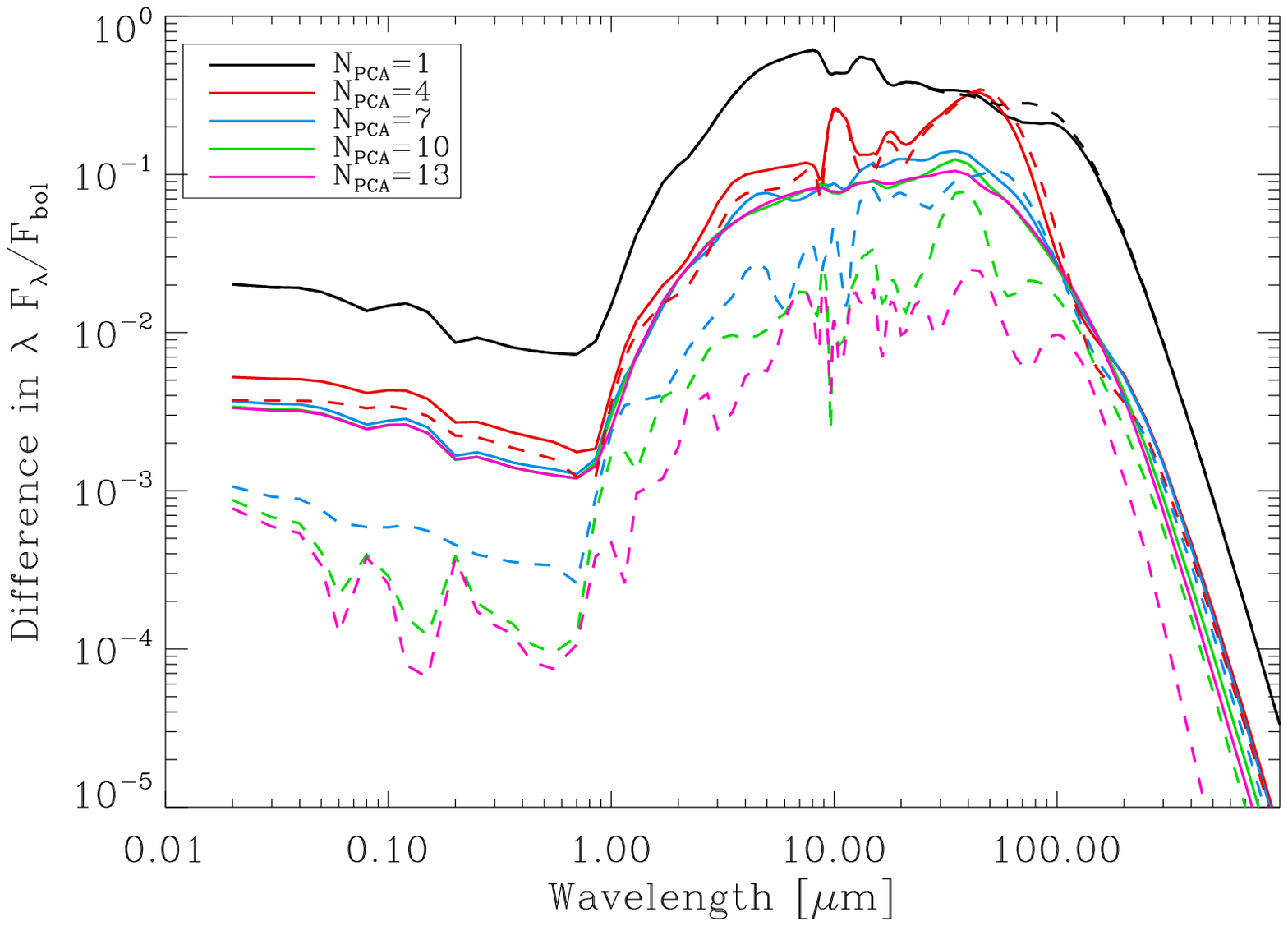}
\caption{Reconstruction errors characterized by the standard deviation (left panel) and 99\%
percentile (right panel) of the difference in $\lambda F_\lambda / F_\mathrm{bol}$ between all the SEDs of the 
original database and the reconstructed SEDs using an increasing number of PCA eigenvectors
for the reconstruction. The quantity $F_\mathrm{bol}$ is the bolometric luminosity of the AGN.
The dashed lines show the results obtained when the PCA coefficient
of each SED is obtained using the exact SED for projecting along each eigenvector. The
solid line corresponds to the results obtained when the ANN is used to obtain the projection along
each PCA eigenvector. Note that the quality of the reconstruction depends on wavelength 
due to the existence of variable features in some spectral ranges.}
\label{fig:error}
\end{figure*}
\begin{figure*}
\plotone{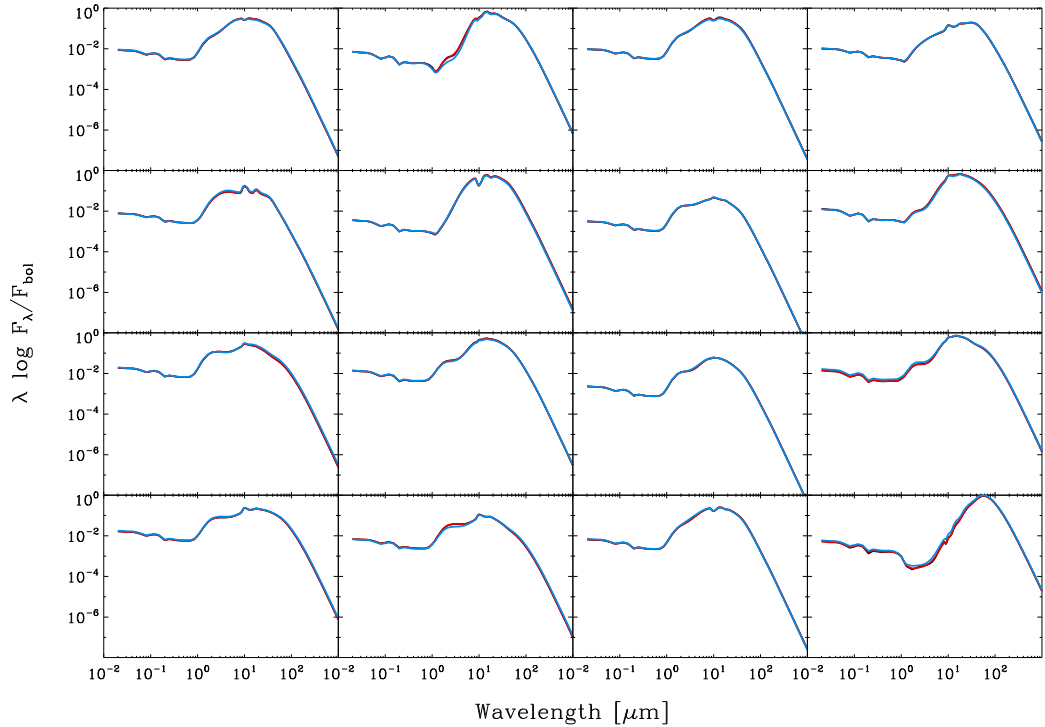}
\caption{Example of the interpolation obtained with the ANN approach. The black lines are the SEDs 
corresponding to randomly chosen physical parameters. The red lines are those reconstructed using the first
13 PCA eigenvectors using the projection of the exact SED on each PCA eigenvector. The blue lines are the
results obtained using the ANN-based reconstruction. Note that the real SEDs are not
seen because the rest of curves overlap.}
\label{fig:SED_reconstruction}
\end{figure*}

\subsection{Neural Network}
Although the PCA dimensionality reduction step has reduced the size of the database, it is still
complex and time consuming to obtain the SED for values of the parameters not present in
the original grid. For this reason, we have developed an interpolation method based on an 
artificial feed-forward neural network \citep[ANN; see e.g.,][]{neal93}, a widespread machine learning 
technique that usually presents very good behavior.
We have developed $N'$ simple neural networks whose topology is shown in 
Fig. \ref{fig:neural_network}. The ANN consists of an input layer formed by 6 neurons that accept 
the physical parameters of the CLUMPY models. The output layer is formed by one neuron whose
value is the projection of the SED corresponding to the physical parameters of the input layer
along each PCA eigenvector calculated before. Both layers are fully connected by an intermediate
hidden layer. The approximation properties of three-layered neural networks is something known after
the universal approximation theorem \cite[e.g.,][]{cybenko88,neal93}. This theorem states that such
a neural network, with a sufficiently large number of neurons in the hidden layer, can approximate any 
continuous function. We have verified that, in 
order to get a compromise between the approximation abilities of the neural network and
the speed of evaluation, values of $N_h$ between 30 and 50 give very good results.
This method is not optimal because the values of $N_h$ are set
empirically. More refined methods probably based on Bayesian model selection \citep[or the
approximate minimum description length method used by][]{asensio_ramos06} can be used
to infer the optimal number of hidden neurons based on objective measurements. However, for the
purpose of our work, the employed method is enough to ensure good approximation properties 
while maintaining a fast execution speed.
The ANN uses the hyperbolic tangent activation function. Prior to utilizing
the neural network, all the input and output values are normalized to the interval
$[-1,1]$ to improve the interpolation abilities of the network.

The training of the ANN is done by modifying the weights until the minimizing the quadratic
difference between the output of the neural network and the correct values of the
database (see appendix \ref{sec:appendix_ann} for more details). It is important to note that 
over-fitting has been avoided using two data-sets chosen randomly 
from the database: one for training and one for validation purposes. The training process is stopped when the
quadratic error decreases for the training set but starts to increase for the validation set.
We have verified that it is possible to carry out the training of the neural networks
using only a subset of the full database, which greatly accelerates the process. This is a consequence of the 
smooth variation of the SEDs with the physical parameters. Picking randomly from the
database a training set with $\sim 10$\% of the total number of models, the trained neural 
network does a very good job with the validation set and with the whole database.

In order to analyze the ability of the PCA+ANN combination to reproduce the database, we show in 
Fig. \ref{fig:error} the standard deviation (left panel) and the 99\% percentile (right panel) of the distribution 
of differences in $\lambda F_\lambda/F_\mathrm{bol}$ between
the exact SEDs of the full database and the reconstructed SEDs. The 99\% percentile has been
also represented in order to test the possibly poor generalization properties of neural networks in regions
close to the boundaries of the space of parameters.
The dashed lines present the results
when the PCA reconstruction is done using the original database. In such a
case, the reconstruction error is monotonically decreasing with the number of included PCA
eigenvectors. If all the eigenvectors are used, the reconstruction error turns out to be identically zero. With the
first 13 eigenvectors, the reconstruction errors are below 5$\times$10$^{-3}$ ($1\sigma$) and 10$^{-2}$ (99 \%
percentile) in all the wavelength range of 
interest. The solid lines are the reconstruction errors when the projections along the
PCA eigenvectors are calculated by evaluating the artificial
neural networks. Obviously, due to the approximate character of the neural networks' interpolation, the
reconstruction errors are larger than in the exact case. The approximation abilities of the neural
networks worsen when the order of the eigenvector increases. The reason is that these eigenvectors
contain high-frequency or less abundant signatures whose variation with the parameters are 
less smooth. However, standard deviations of the reconstruction error are below 2$\times$10$^{-2}$ in the whole 
spectral domain of interest, with errors going down to 10$^{-3}$ in some spectral windows. Concerning the
99\% percentile, errors are always below 10$^{-1}$, with wavelength regions close to 10$^{-3}$.

An example of the ability of the ANN to approximate the
database is shown in Fig. \ref{fig:SED_reconstruction}, where the black line is the exact SED obtained
from the database (note that although only part of the database was used in the training, these SEDs
are obtained randomly from the full database). The red line is the SED reconstructed using only the
first 13 eigenvectors but using the correct SED for the projections along the eigenvectors. The 
blue line is the SED reconstructed using the neural networks. Note that differences are hardly
noticeable and are well below any possible observational error or indeterminacy in the physical
properties of the AGN.

\begin{table}
\begin{center}
\caption{Centaurus A high spatial resolution nuclear density measurements from
\cite{Meisenheimer07} and \cite{Radomski08}.\label{tab:fluxes}}
\begin{tabular}{lcc}
\tableline
\tableline
Filter & Central wavelength (\micron) & Flux Density (mJy)  \\
\tableline
NACO J     & 1.28  & 1.3$\pm$0.1     \\
NACO H     & 1.67  & 4.5$\pm$0.3     \\
NACO Ks    & 2.15  & 33.7$\pm$2.0    \\
NACO L'    & 3.80  & 200$\pm$40      \\
T-ReCS Si2 & 8.74  & 710$\pm$40      \\
T-ReCS Qa  & 18.3  & 2630$\pm$650    \\
\tableline

\end{tabular}
\end{center}
\end{table}

\subsection{Advantages}
There are two main advantages of the approach followed in this paper. On the one hand, it opens the
possibility to interpolate in the database, so that it is now possible to calculate SEDs for 
combinations of parameters that were not present in the original grid. In principle, if the 
SEDs depend smoothly on the parameters, the neural networks should have captured all the
variability and there is no necessity to improve the griding of the original database.
On the other hand, the synthesis of the SED is extremely simple and fast because all the
details of the calculations inherent to the CLUMPY model are approximated by the neural networks.
One has to calculate the 13 projections of the SED onto the PCA eigenvectors by evaluating the 
13 neural networks given by Eq. (\ref{eq:neural_network}).
Then, the reconstruction of the SED is obtained by adding the first 13 PCA eigenvectors weighted by
these 13 projections. In terms of computational time, this makes it possible to synthesize
$\sim 10^4$ SEDs in just a minute, so that we obtain a gain in time of a factor 10$^3$ or larger. 
The ANN+PCA approach can be essentially considered as a huge compression of the database. Instead
of saving the whole database, one only needs to save the weights of the neural networks and the PCA eigenvectors.
In our case, the complete database amounts to $\sim$430 Mb, while the ANN+PCA approach amounts to $\sim$40 kb, with
a reduction factor that is close to 10$^4$, with the added benefit of being able to easily interpolate.
Of course, this improvement in the calculation speed is compensated by small differences as compared with
the correct SEDs.

\subsection{Simulating filter photometry}
For the cases in which the observations are of filter photometry kind, 
once the SED is obtained with the previous formalism, it remains to simulate the 
effect of the filters on the simulated SED.
Given that $\phi(\lambda)$ is a filter normalized to unit area used to obtain a point in the observations, the 
synthetic value is obtained by just evaluating:
\begin{equation}
f(\lambda_c) = \int_{-\infty}^{\infty} F(\lambda) \phi(\lambda) \mathrm{d}\lambda,
\end{equation}
where $\lambda_c$ is the central wavelength of the filter, obtained as:
\begin{equation}
\lambda_c = \int_{-\infty}^{\infty} \lambda \phi(\lambda) \mathrm{d}\lambda.
\end{equation}
Both integrals are approximated in \B\ with a very simple trapezoidal quadrature:
\begin{equation}
\tilde f(\lambda_c) = \sum_i w_i F(\lambda_i) \phi(\lambda_i),
\end{equation}
with $w_i$ the weights of the trapezoidal quadrature.

\section{Illustrative example}
We have implemented this forward modeling code into a Markov Chain Monte Carlo sampling algorithm that is
used to evaluate the posterior distribution function for the physical parameters once an observation
is provided (see appendix \ref{sec:bayesian_inference} for details). The filters at 
which the information is available are fully configurable from a
database of filters belonging to different instruments. One only needs to
select the filter and give the the observed flux, $d_i$, and its corresponding error.
For the moment, Gaussian errors with standard deviation $\sigma_i$ and upper limits within a 
certain (user-selectable) confidence level are possible.

The simplest version of the code uses uniform priors for all the parameters and admit to give the upper and
lower limits for every parameter. It is also possible to choose a Gaussian prior 
in case any parameter is known a-priori to be around a given value with a certain dispersion. The central position
and the width of the Gaussian are the only adjustable hyper-parameters\footnote{The name hyper-parameters are
usually applied to describe parameters that modify the prior distribution. We treat these hyper-parameters
as fixed and adjustable by the user and we do not carry out any estimation over them.} in this case.
More complex prior distributions are straightforward to include in the
code and can be fully configurable. 

Apart from the 6-dimensional vector of parameters of the CLUMPY model given by
$(\sigma,Y,N,q,\tau_V,i)$, we add other parameters like a vertical shift in logarithmic scale
that accounts for the normalization in luminosity of the SED or the amount of interstellar extinction
characterized by an extinction law and the absorption at visible wavelengths.
We neglect the presence of extinction in this work, but the effect of the normalization in
luminosity is treated as a \emph{nuisance} parameter\footnote{Parameters in which
the model depends on but are of no interest in the parameter estimation.} and all the results are
presented with this parameter integrated out (marginalized). Since the marginal posterior distributions of the vertical shift are
already an output of \B, one can carry out inference over the total luminosity of the AGN.

\subsection{Observations}
High resolution infrared data have been compiled from the literature for the 
Seyfert 2 galaxy Centaurus A (NGC~5128) in order to construct a purely-nuclear 
SED. Centaurus A is the closest active galaxy, with its core heavily obscured by
a dust lane, and consequently only visible at wavelengths longwards of 0.8
\micron~\citep{Schreier98,Marconi00}.

\begin{figure*}
\includegraphics[width=0.32\textwidth]{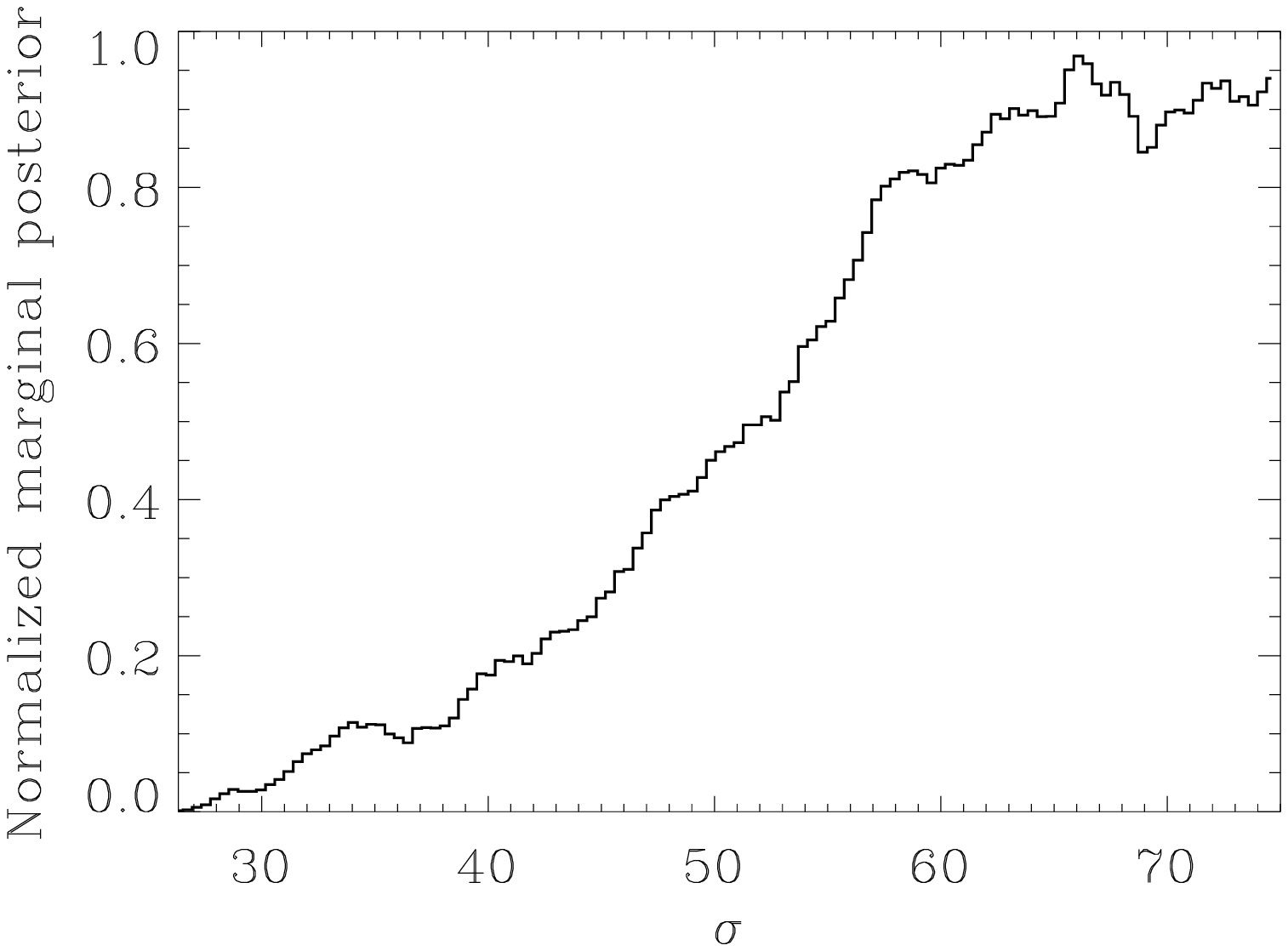}
\includegraphics[width=0.32\textwidth]{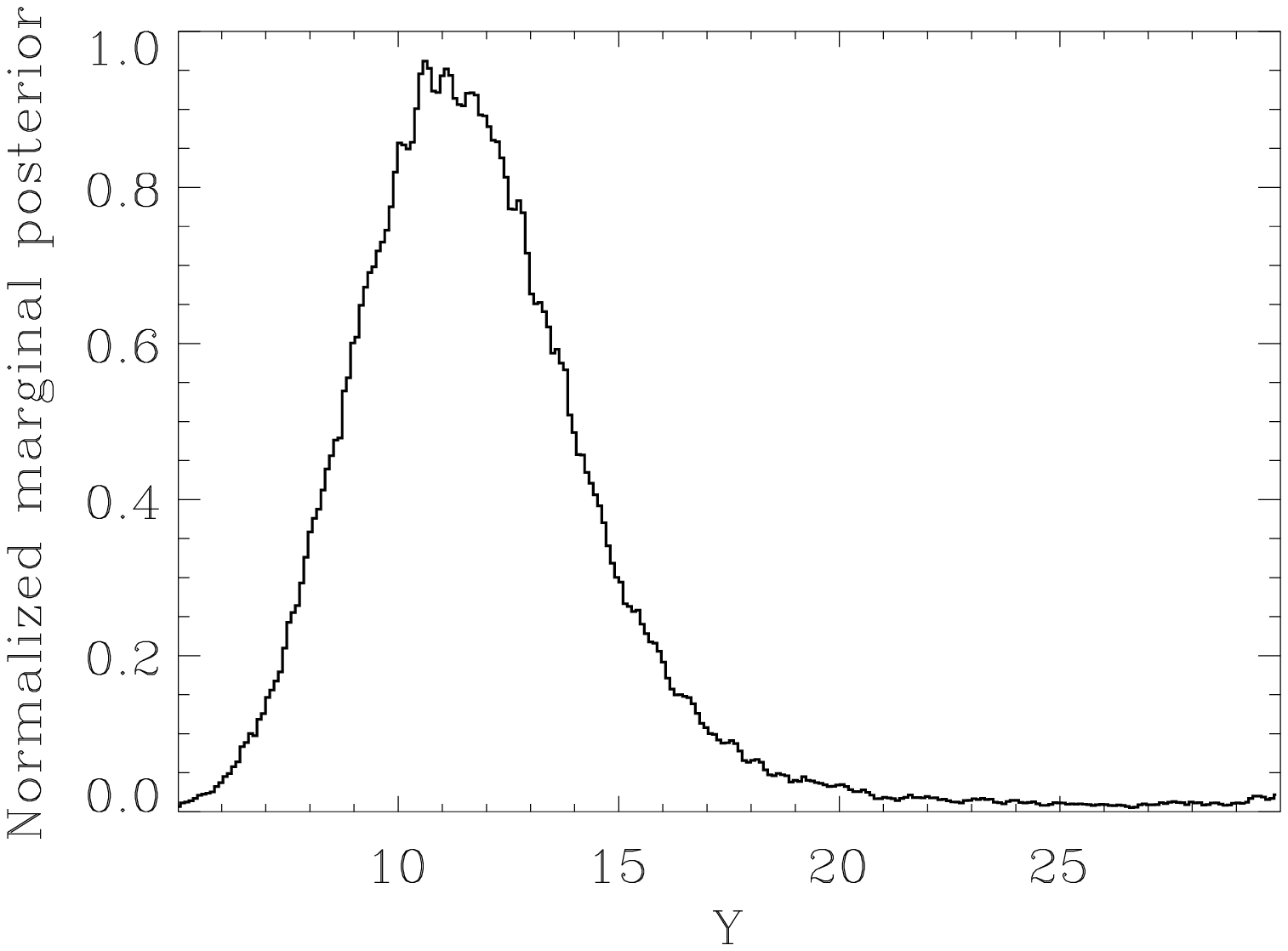}
\includegraphics[width=0.32\textwidth]{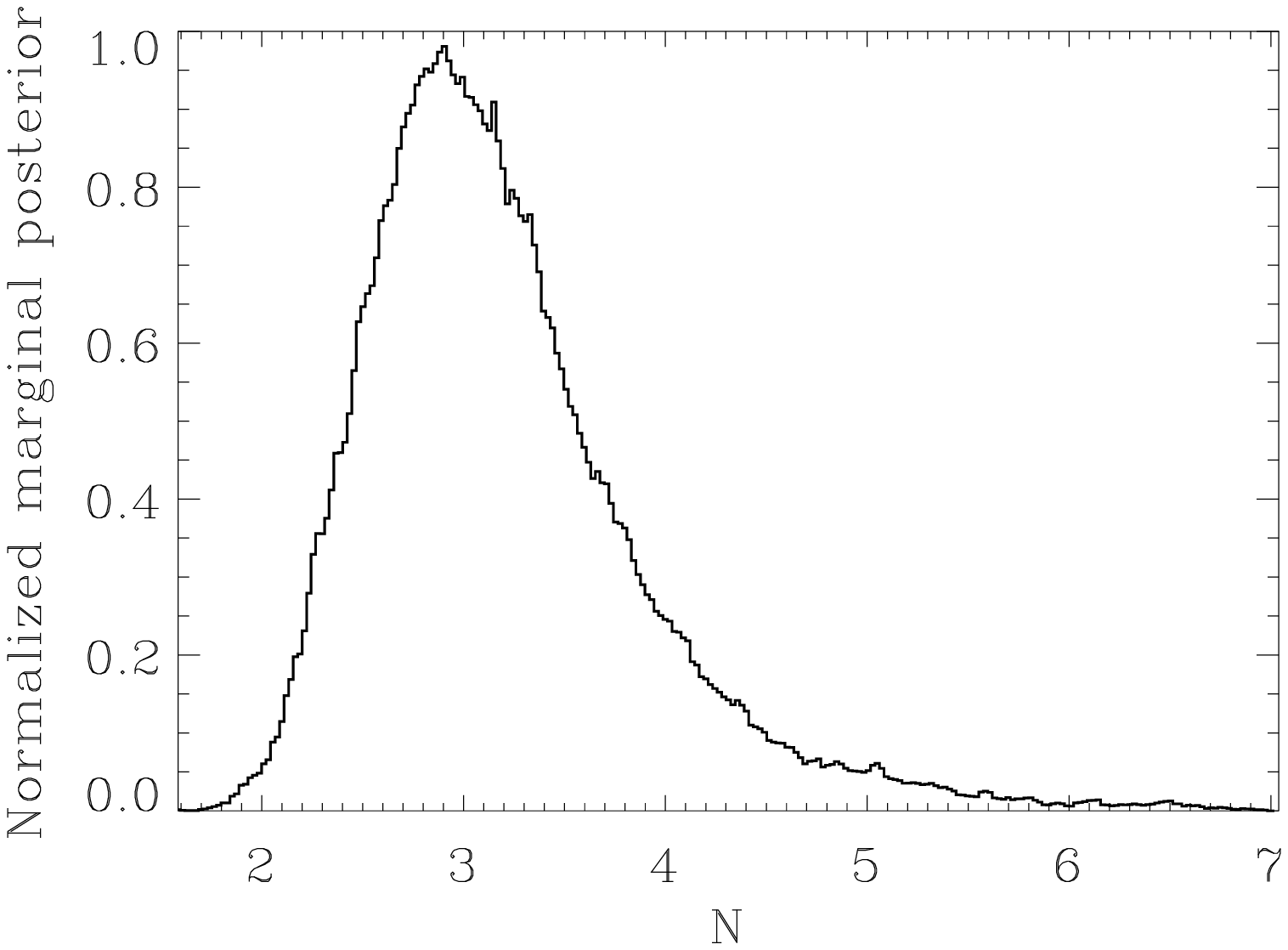}
\includegraphics[width=0.32\textwidth]{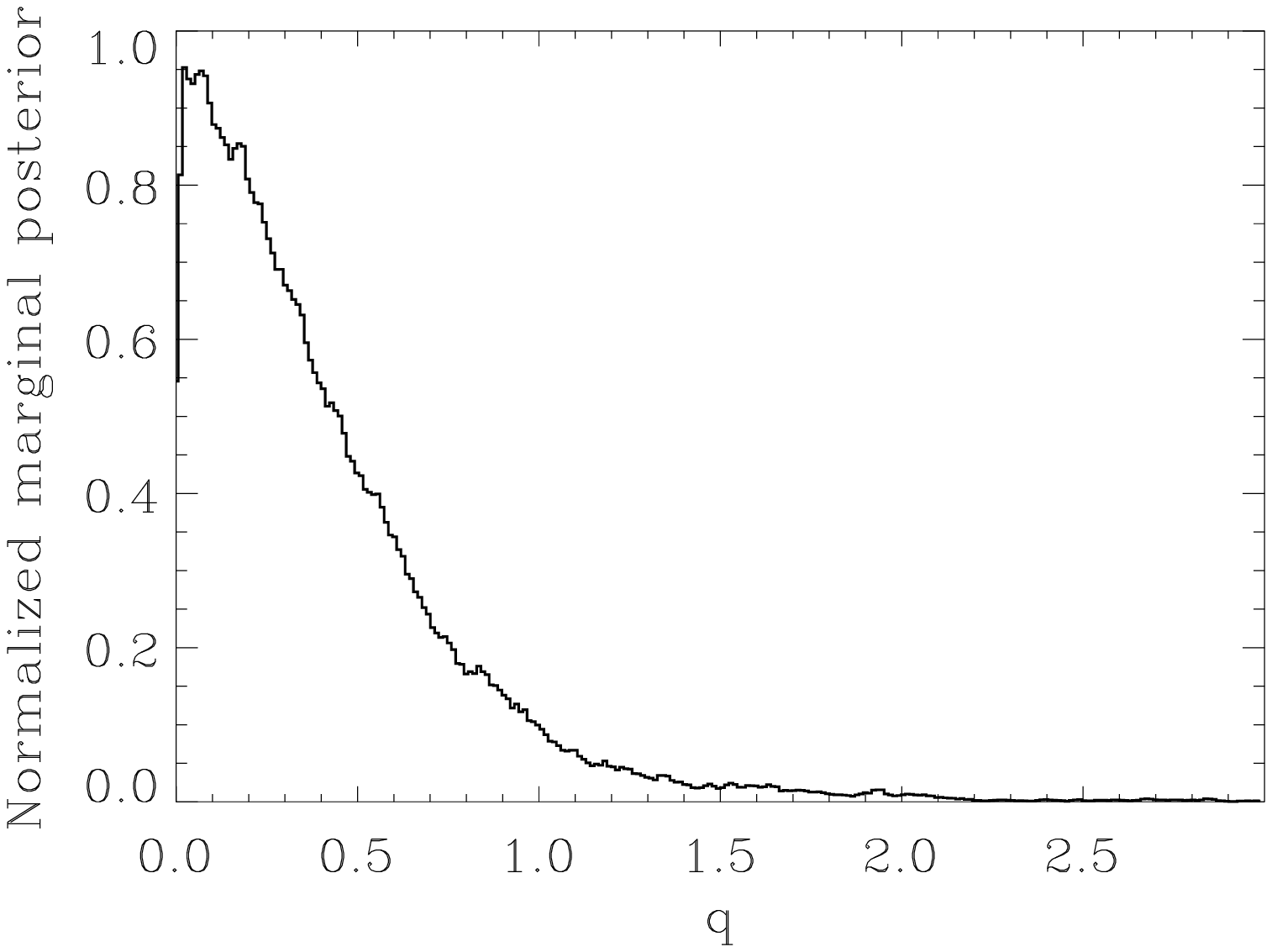}
\includegraphics[width=0.32\textwidth]{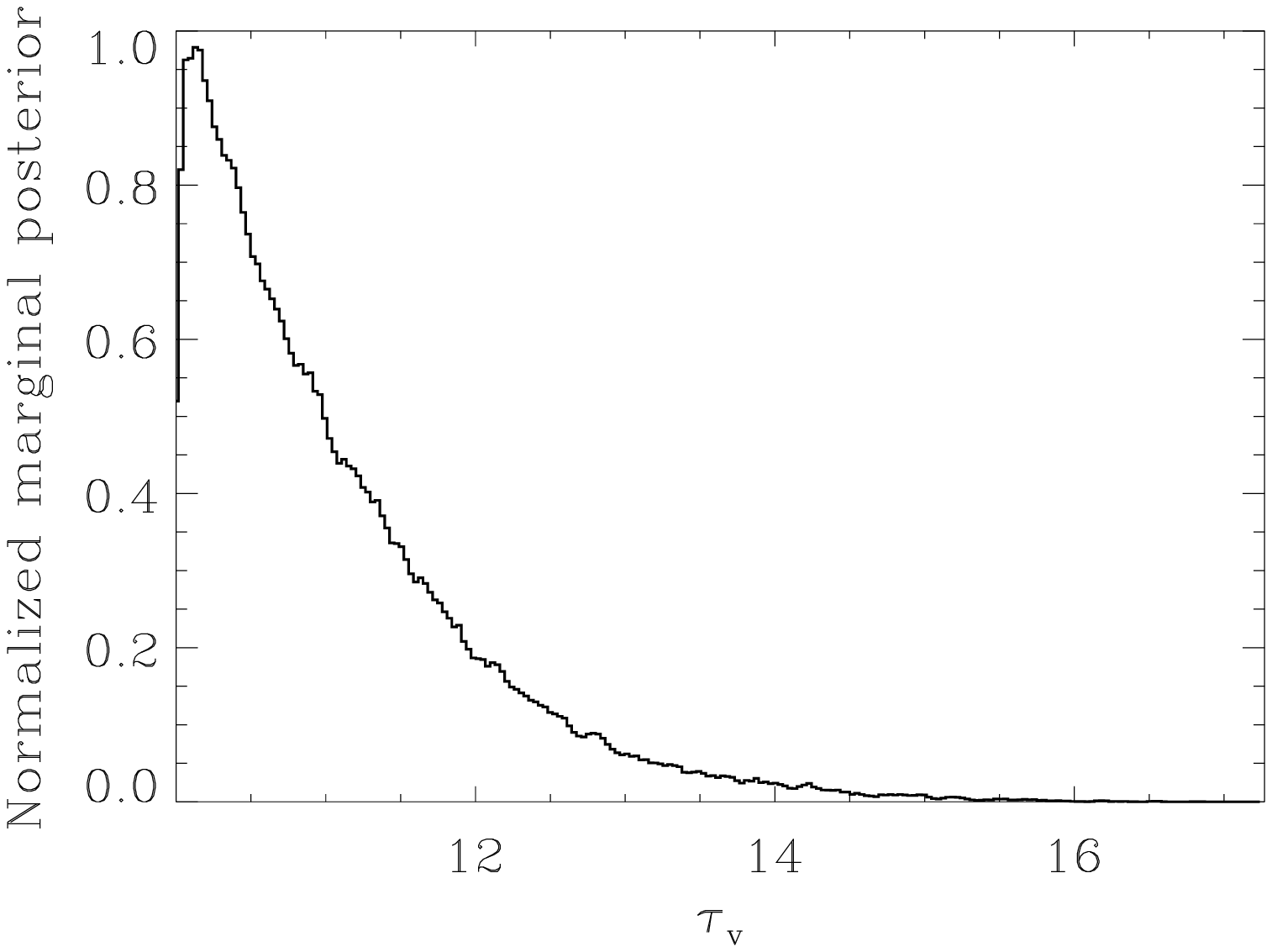}
\includegraphics[width=0.32\textwidth]{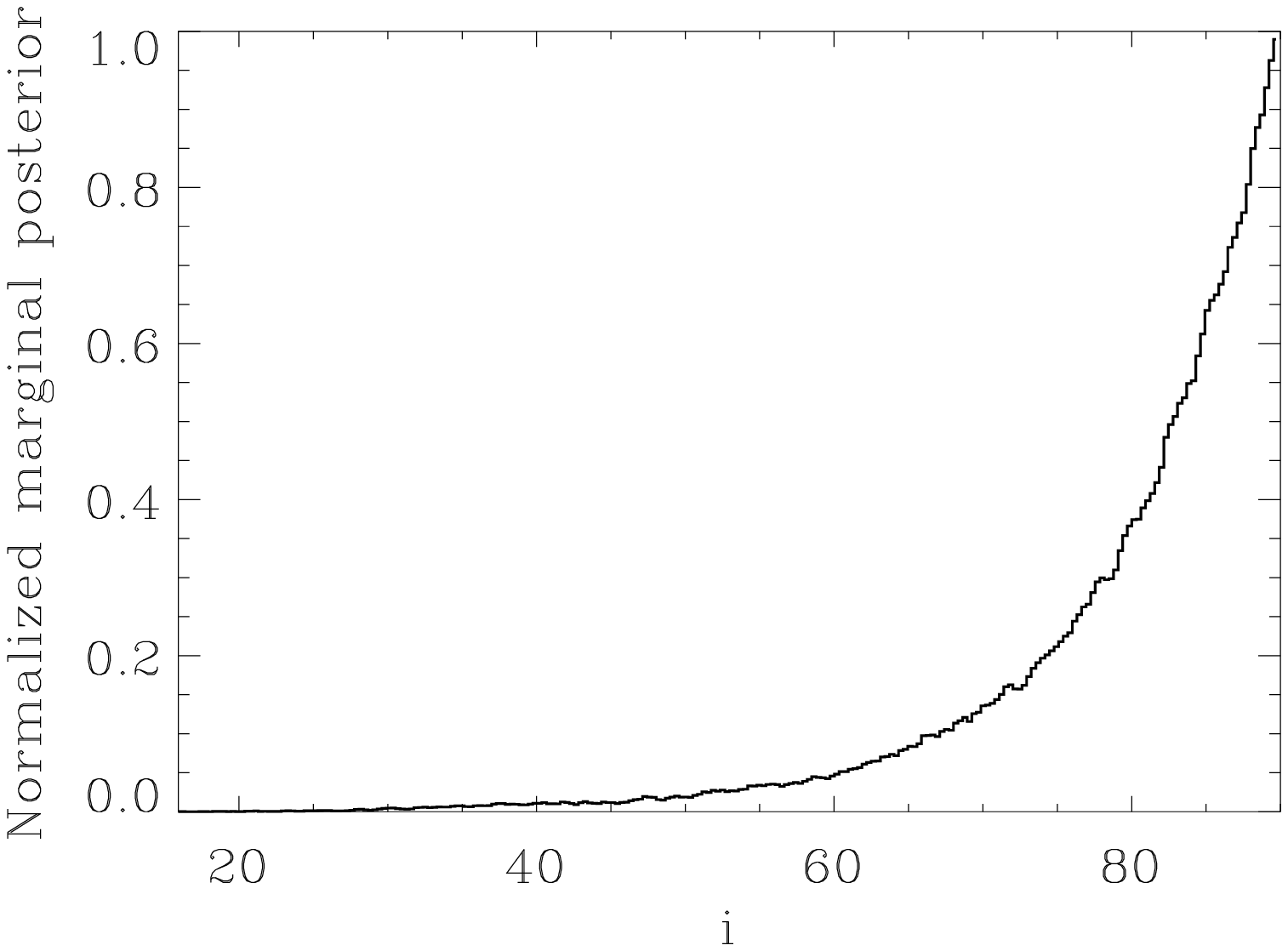}
\caption{Marginal posterior distributions for all the free parameters of the CLUMPY model considered here. Note that $\tau_V$ and
$N$ are nicely constrained. Other distributions only favor certain values of the space of parameters.
The vertical shift of the SED, treated as a nuisance parameter, has been marginalized.}
\label{fig:marginal1D_cenA}
\end{figure*}

Near-infrared data obtained with Naos-Conica (NACO) at UT4 have been compiled  
from \cite{Meisenheimer07}. 
Conica is a high spatial resolution near-infrared imager and spectrograph
\citep{Lenzen98}, that works together with Naos, the Nasmyth Adaptive 
Optics System \citep{Rousset98}, providing adaptive-optics corrected
observations in the range 1-5 \micron.
\cite{Meisenheimer07} found the nucleus unresolved at all wavelengths with a
FWHM of 0.10\arcsec\ in the J band, 0.088\arcsec\ in the H band, 0.059\arcsec\
in the Ks band, and 0.090\arcsec in the L' band. 
The fluxes corresponding to this unresolved component 
have been compiled and reported in Table \ref{tab:fluxes}.

High spatial resolution mid-infrared data of the nucleus of Centaurus A are also compiled from
\cite{Radomski08}. The observations were taken in the Si2 filter (8.8 \micron)
and in the Qa band (18.3 \micron) using the mid-infrared imager/spectrometer
T-ReCS on Gemini South \citep{Telesco98}. 
The core is also unresolved in this range, surrounded by a diffuse extended 
emission. In these bands, the upper limits to the size of the unresolved nucleus are
0.19\arcsec\ at 8.8 \micron\ and 0.21\arcsec\ at 18.3 \micron\ at the FWHM level.

With these high spatial resolution density measurements of the unresolved component of 
Centaurus A, we can construct a purely-nuclear SED of this galaxy, to be fitted
with the CLUMPY models, as an example of the use of our tool. 

\subsection{Bayesian Analysis}
The parameter estimation results presented in this section have been carried out with a
Markov Chain of length $6 \times 10^5$, of which we take out the initial 40\% as a burn-in \citep[transitory
initial portion of the Markov Chain in which the algorithm is not correctly sampling the
distribution; see, e.g.,][]{neal93,gregory05}.
Our experiments have shown that the transitory phase is quite short and that the burn-in 
can be safely reduced to just the first $\sim 10$\% of the chain without much problem, thus increasing the 
quality of the sampling. Although a 40\% burn-in is surely excessive, the fact that the
MCMC scheme works very fast (less than half a minute for a $6 \times 10^5$ chain), we prefer 
to run a longer chain and maintain the large burn-in.
This way we make sure that the MCMC algorithm is well mixed and we are sampling correctly the posterior
distribution.

The one-dimensional marginalized posterior distributions are shown in Fig. \ref{fig:marginal1D_cenA} for
all the free model parameters. These distributions are obtained as histograms of the Markov Chain for
each parameter due to the automatic marginalization properties of the MCMC technique.
This information encodes, for every parameter, the effect of ambiguities and degeneracies
(the marginalization of the posterior takes into account all the possible values of the rest of parameters
weighted by their probabilities) and summarizes the statistical properties of the estimation for
each parameter. Uniform priors have been employed, leaving for a later publication the analysis of AGN in which a-priori
knowledge might be present (Ramos Almeida et al., 2009, in preparation). Therefore, these posterior 
distributions have to be compared with uniform 
distributions giving equal probability to all combinations of parameters. When the observed data 
introduces enough information into the problem, the posterior distributions clearly differ from
the uniform distribution. The uniform distributions are truncated to the following intervals:
$\sigma=[15,75]$, $Y=[5,30]$, $N=[1,15]$, $q=[0,3]$, $\tau_V=[10,60]$ and $i=[0,90]$. These
values are based on physically reasonable assumptions that avoid non-realistic solutions.

Figure \ref{fig:marginal1D_cenA} clearly indicates that all the parameters are 
nicely constrained for Centaurus A. The marginal distribution of $\tau_V$, with its asymmetric shape, is showing us that
the observations are able to put an upper limit to its value. The calculation of the confidence intervals in
this histogram gives us that the upper limit to $\tau_V$ is 11.2 at 68\% ($1\sigma$ level) confidence and 13.0 at
95\% confidence ($2\sigma$ level). Concerning $N$, there is a tail towards large values that produce
slightly asymmetric error bars. If we summarize the histogram using the median, we find that $N=3.1^{+0.7}_{-0.5}$
at 68\% confidence and $N=3.1^{+2.0}_{-0.9}$ at 95\% confidence. It is also possible to use different
quantities to summarize the statistical information, like the posterior mean or the posterior mode 
(maximum-a-posteriori or MAP), although
the relevant information is provided by the full marginal posterior distribution.


\begin{figure*}
\includegraphics[width=0.32\textwidth]{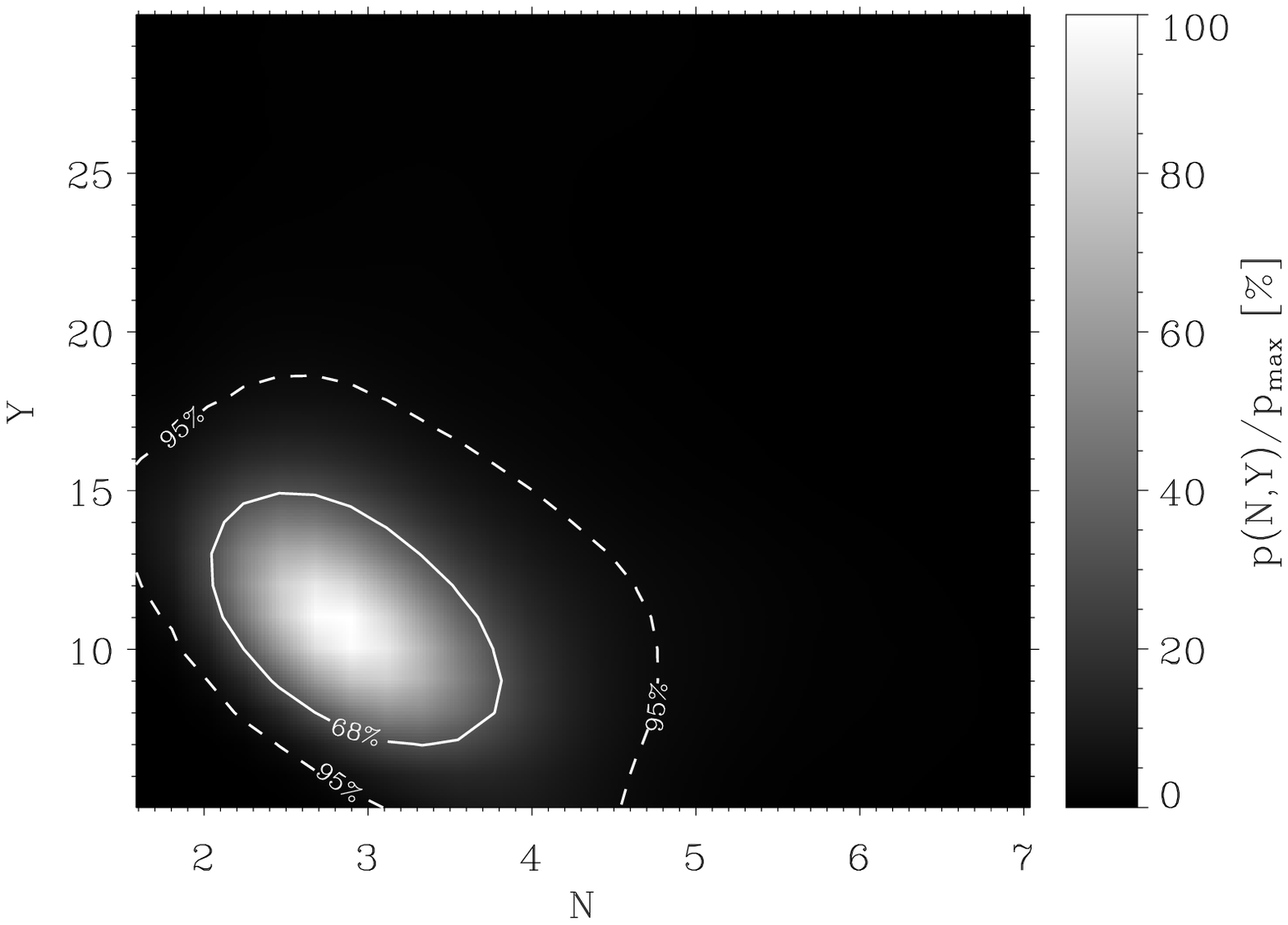}
\includegraphics[width=0.32\textwidth]{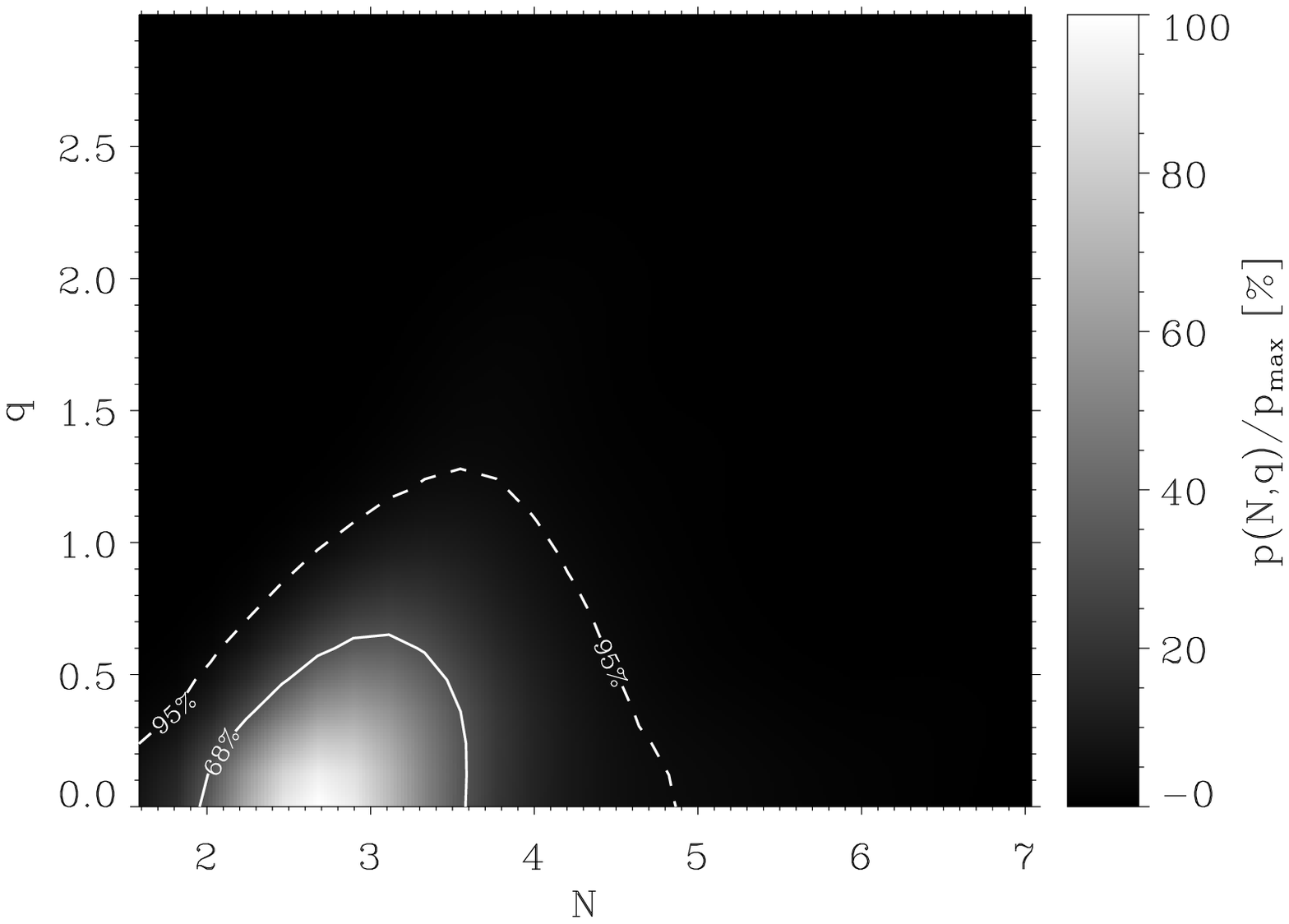}
\includegraphics[width=0.32\textwidth]{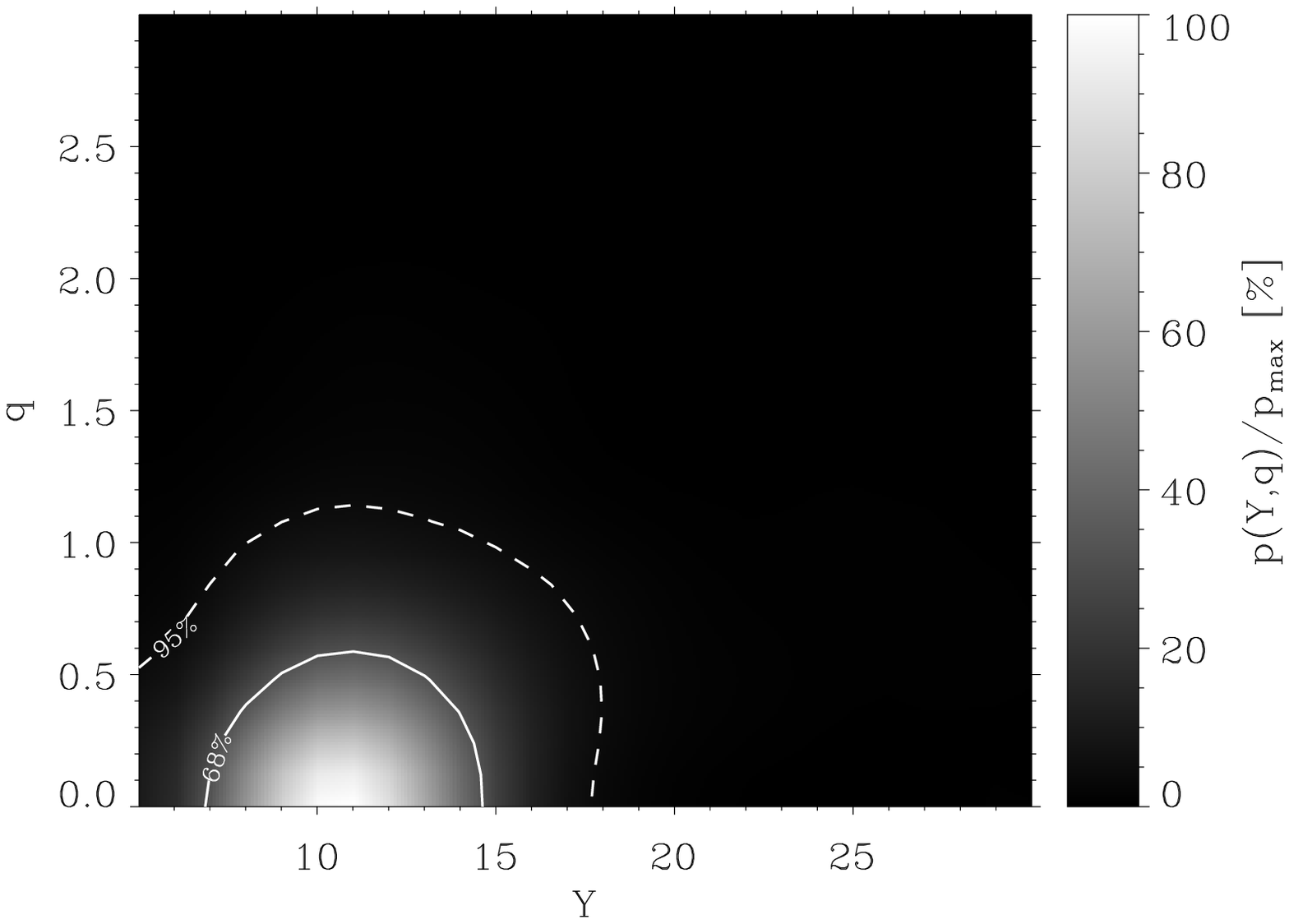}
\includegraphics[width=0.32\textwidth]{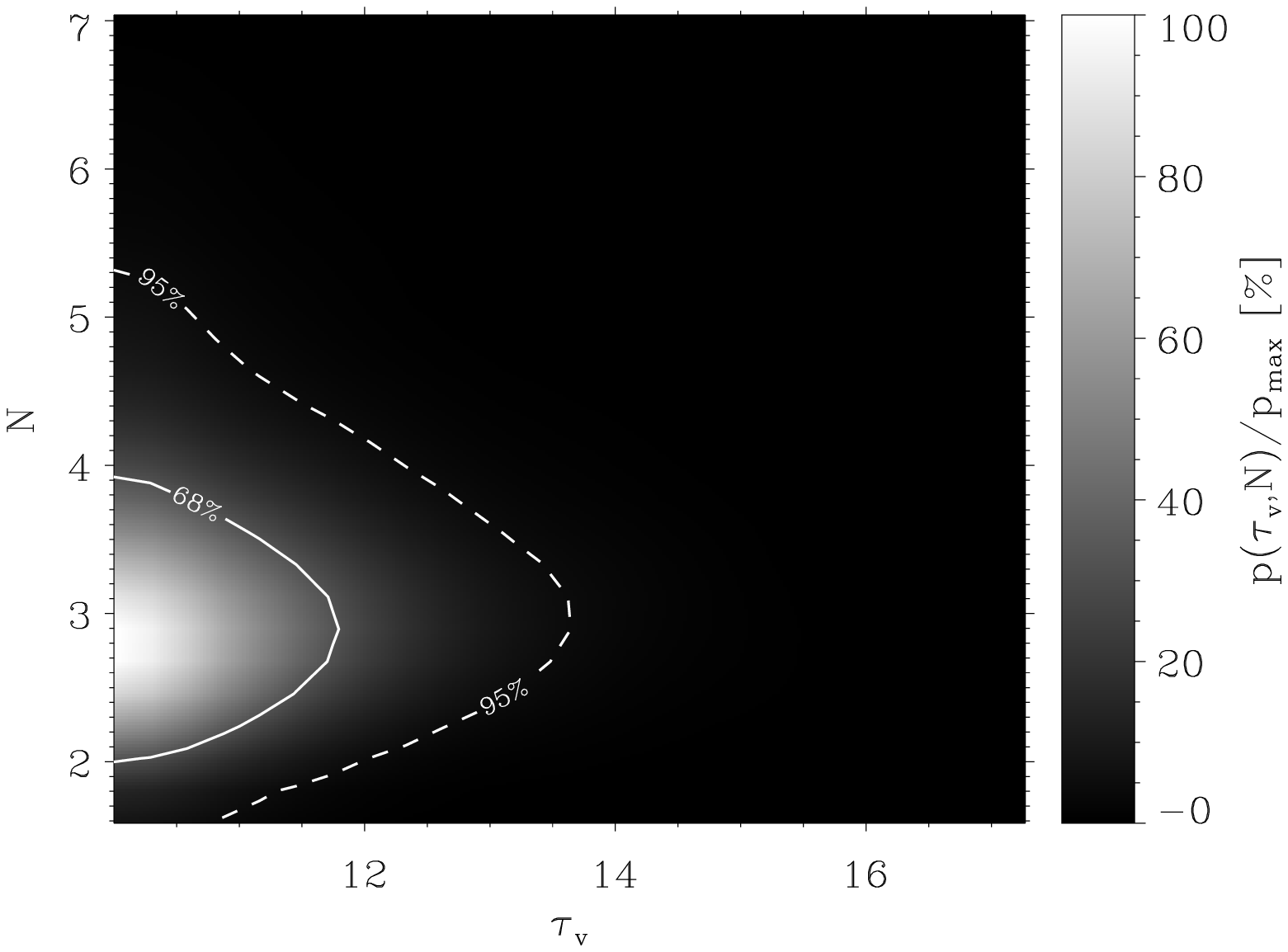}
\includegraphics[width=0.32\textwidth]{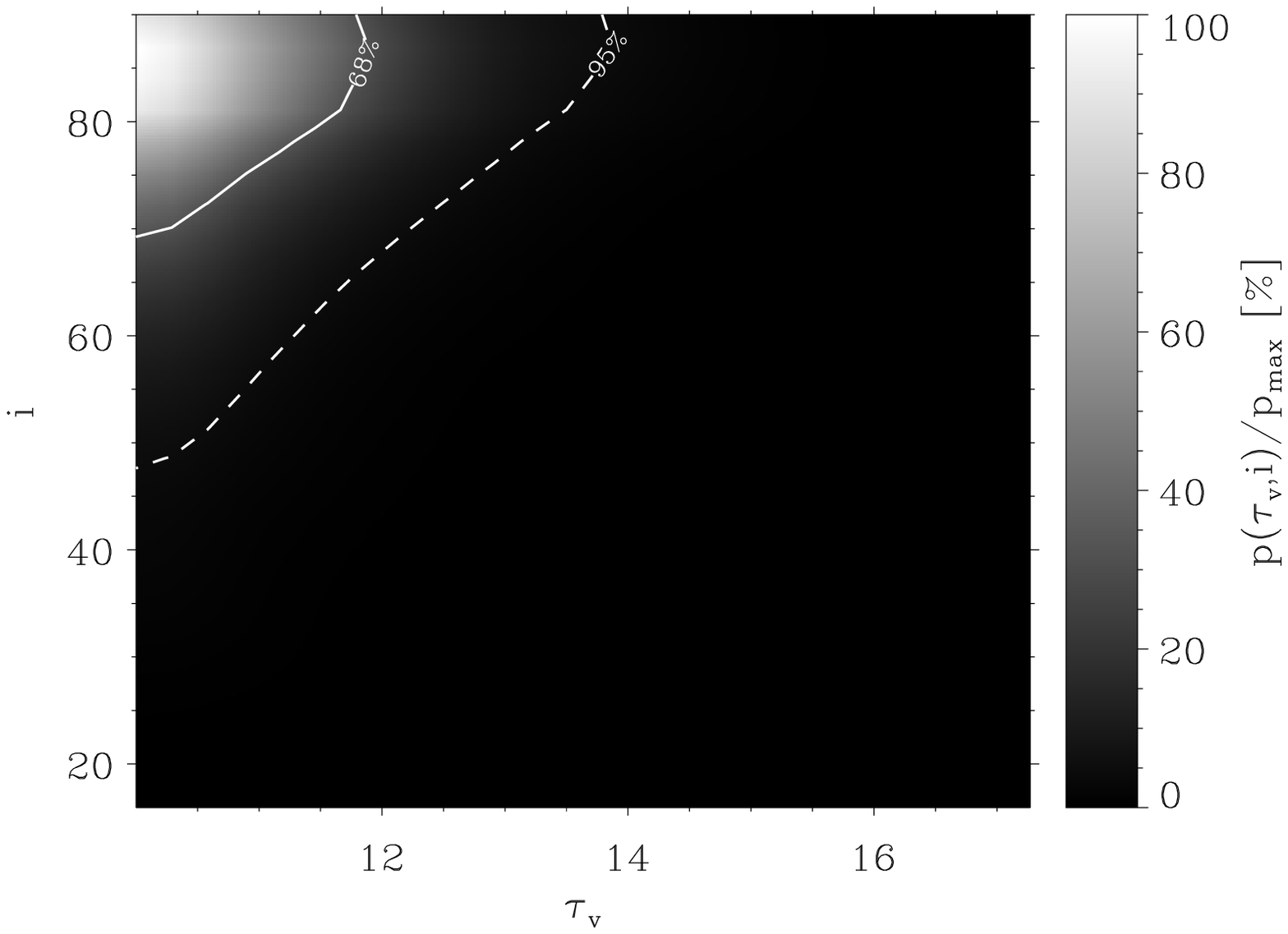}
\includegraphics[width=0.32\textwidth]{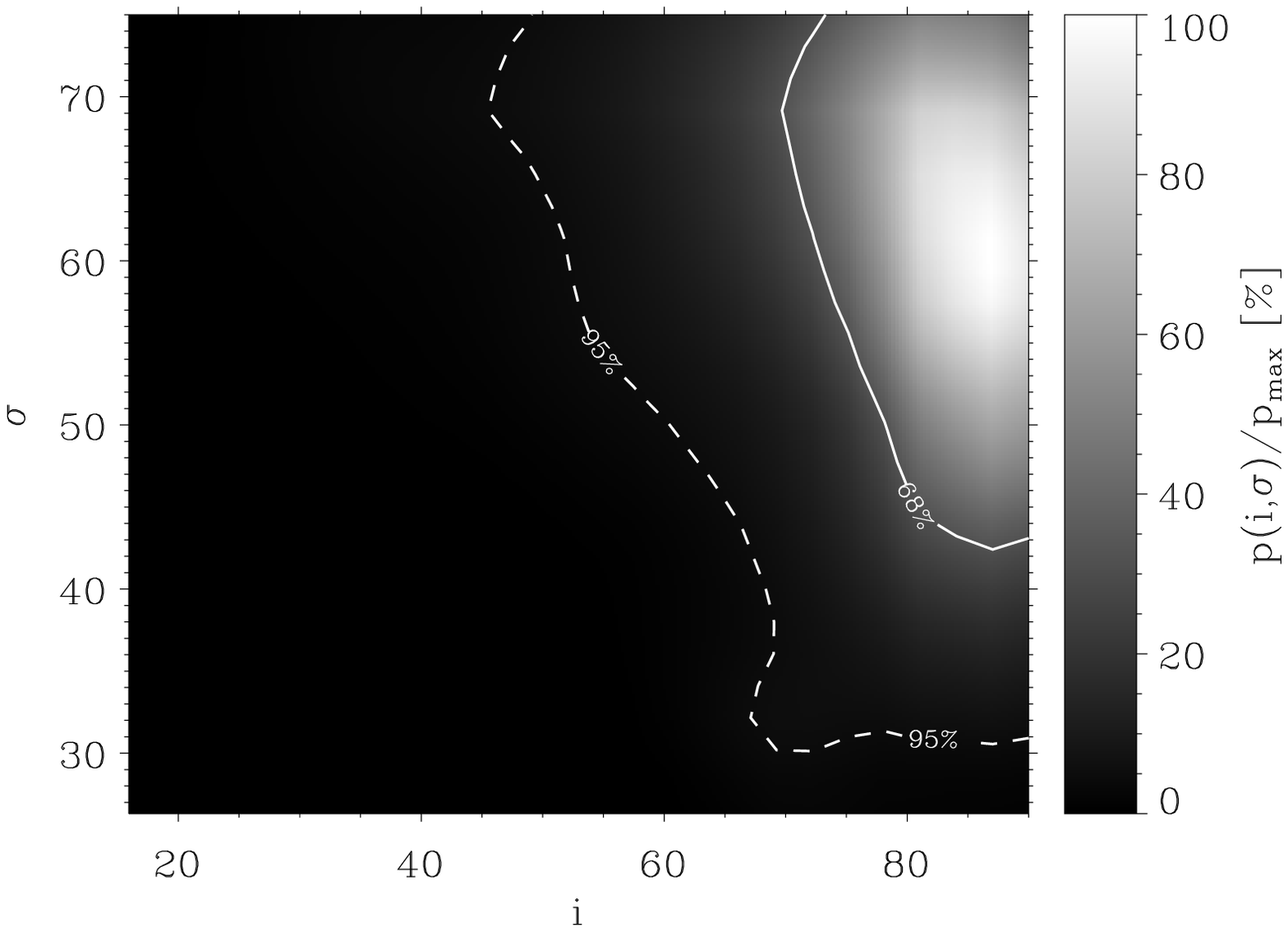}
\caption{Two-dimensional marginal posterior distributions (joint distribution) for several combinations of parameters.
The contours indicate confidence regions at 68\% and 95\%. 
}
\label{fig:marginal2D_cenA}
\end{figure*}

The marginal posterior for $Y$ also indicates that it is a nicely constrained parameter, again with
asymmetric error bars due to the enhanced tail. We obtain $Y=11.5^{+2.9}_{-2.3}$ at 68\% confidence
and $Y=11.5^{+8.6}_{-4.3}$ at 95\% confidence. A different behavior is found for $\sigma$, $q$ and $i$. In the three
cases, there is a region of the space of parameters that is favored with respect to others. For instance,
the Bayesian analysis gives a larger probability to inclinations close to 90$^\circ$ (the MAP value is
$\sim 90^\circ$), practically discarding angles close to 0$^\circ$. The same happens for $q$, in
which data favor values close to zero. Summarizing, we get $i>78^\circ$, $\sigma>56^\circ$
and $q < 0.47$ with 68\% confidence.


Correlations between the parameters can be understood at the light of the two-dimensional marginal
distributions of Fig. \ref{fig:marginal2D_cenA}, where the contours mark
the 68\% and 95\% confidence regions. Instead of plotting all possible combinations
of parameters, we only show six cases that are representative of the general behavior. 
It can be seen that the observed data discard large values of $N$, $Y$, $\tau_V$ and $q$. 
The plot $\sigma-i$ presents weakly constrained parameters, although it is clear from this 
figure that large values of $i$ are favored (in 
accordance with the Type-2 classification of Cen A). Small values of $i$ are 
not preferred and become slightly more likely as $\sigma$ increases (i.e., as the probability
that the central engine is blocked from view). In other words, the results show that Type-1
orientations (small $i$ simultaneous with small $\sigma$) are not favored by the data. We present
the physical interpretation of these results in Ramos Almeida et al. (2009, in preparation).


It turns out interesting that there is not much ambiguity in the parameters so that 
the available data is able to constrain the large variability of the clumpy torus models. One should
expect to obtain even more constrained marginal posterior distributions when increasing the number of filters.

\begin{figure}
\includegraphics[width=0.7\columnwidth,angle=90]{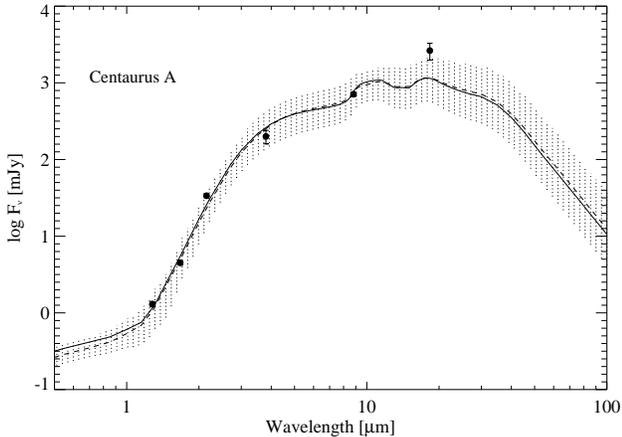}
\caption{High-resolution nuclear SED of Centaurus A, constructed with near- and mid-infrared data shown
in Table \ref{tab:fluxes}. The solid line
presents the maximum-a-posteriori estimation of the SED while the dashed line is the one calculated
with the median of the marginal posterior distributions of the parameters. The shaded region is the range of
SEDs compatible within the 68\% confidence interval for each parameter.}
\label{fig:SED}
\end{figure}

Although the solution to the Bayesian inference problem are the posterior distributions
shown in Fig. \ref{fig:marginal1D_cenA}, one can try to represent the models corresponding
to the median or the maximum-a-posteriori values of the parameters (within the confidence intervals) 
to visually compare them with the observations. The maximum-a-posteriori 
model is displayed in solid line in Fig. \ref{fig:SED}. This
is obtained using the combinations of parameters that maximize the posterior distribution. The
dashed line shows the model obtained using the medians of the marginal posterior distribution for 
each parameter. Finally, the dashed line tries to give an idea of the range of 
variability of the compatible models. It is built by synthesizing SEDs for all combinations
of parameters taking into account their confidence intervals around the median value.

\section{Conclusions}
This paper presents a computer code for the Bayesian analysis of nuclear SEDs of AGN using
CLUMPY models. This approach allows us to obtain the full solution to the inference
problem in terms of posterior probability distributions of the model parameters. These 
probability distributions take into account the a-priori information about the
parameters and the information introduced by the observations. Presently, 
the prior distribution can be selected to be uniform in an interval, Gaussian or Dirac 
delta. According to the observations, the code admits observations corrupted with
Gaussian noise and/or upper limit detections.

The machine learning technique based on the combination of the PCA decomposition
and the application of artificial neural networks for the approximation 
of the database leads to a gain of a factor $10^4$ in disk storage and $10^3$ in execution
time. As a sub-product, it provides the possibility to interpolate in the database, thus it becomes
feasible to generate SEDs for combinations of parameters not present in the original grid. This
characteristic will be of great importance for the analysis of the response function of the
SED in certain spectral ranges of interest to the model parameters (derivatives of the SED 
with respect to the model parameters).

Although the space of parameters is very large, the approximation properties of the method are
very good, giving differences between the exact and the approximate SED with a standard deviation
below 0.1 dex in the spectral window of interest. These errors are clearly below any uncertainty
in the observations, so that the final results will not be dominated by the approximation
step. This good behavior is possible because of two reasons. First, the variation of the SEDs is
notably smooth when the physical parameters are varied. This greatly facilitates the interpolation
properties of the artificial neural networks. Second, the precision needed in the 
approximation is not specially significant and one only needs to train the neural networks
until the differences between the exact SEDs and the reconstructed SEDs are below the
observational errors. 

In order to demonstrate the output of the code, we have shown an example with Centaurus A. The
analysis of the marginal posterior distributions shows that some parameters are nicely determined
by the data, while other parameters remain less constrained. Although one can summarize the
marginal posteriors using modes, medians and/or means, together with confidence intervals, the 
true solution to the problem from the Bayesian point of view are the histograms shown in Fig. 
\ref{fig:marginal1D_cenA}.

Although \B\ is focused towards the analysis of nuclear SEDs of AGN using CLUMPY models, the
core of the method remains completely general and can be applied to a myriad of problems for which
one is interested in fitting an already built database of models to observations. Cases like the
database of synthetic spectra computed for the GAIA project \cite[e.g.,][]{phoenix_gaia05} or
the analysis of the spectral energy distributions of protostars \cite[e.g.,][]{robitaille07}
are examples of potential applications. Apart from the gain in speed in the analysis, one would
be able to carry out a fully Bayesian analysis of the observations, thus opening the possibility
of including prior information and/or carrying out marginalization of nuisance parameters.
Furthermore, Bayesian model selection techniques would facilitate the objective selection of
several competing models for explaining the observables.

\B\ is coded in Fortran 90 with a graphical front-end developed in IDL\footnote{\texttt{http://www.ittvis.com/idl}}.
We offer \B\ to the astrophysical community with the hope that it will help researchers to take
advantage of the CLUMPY models for analyzing observed SEDs. To get a copy, it suffices with making an e-mail request to 
the authors of this paper.

\begin{acknowledgements}
We thank Nancy Levenson, Moshe Elitzur, and Ana P\'erez Garc\'{\i}a for useful comments.
Financial support by the Spanish Ministry of Education and Science through projects AYA2007-63881 and AYA2007-67965-C03-01,
the European Commission through the SOLAIRE network (MTRN-CT-2006-035484) and the Spanish MEC under the Consolider-Ingenio 
2010 Program grant CSD2006-00070: First Science with the GTC (http://www.iac.es/consolider-ingenio-gtc/) is gratefully acknowledged.
We also acknowledge ``Los Piratas'' for inspiring the development of this work.
\end{acknowledgements}

\appendix

\section{Bayesian Inference}
\label{sec:bayesian_inference}

\subsection{Fundamental considerations}
\B\ is a computer code that is focused towards analyzing observed SEDs
using theoretical models and carrying out inference over the parameters of the
theoretical models. It is built under the framework of the Bayesian approach for 
the inference \citep[see e.g.,][]{neal93,gregory05},
of which we briefly summarize the main ideas. Let $M$ be a model that is proposed to explain an observed data set $D$.
In our case, $D$ is usually the set of points observed using filter photometry, but more 
information can be introduced if the observation is of spectroscopic type.
Additionally, $M$ is the CLUMPY model described by \cite{Nenkova08a,Nenkova08b} that we
have briefly described in \S\ref{sec:clumpy_models}. Let
$I$ be a set of sensible background a-priori knowledge about the problem (for instance, the filters
used for the observations). In general, the model $M$ is described by a set of equations
or algorithms that depend on a vector of parameters, $\thetabold$. These parameters usually
have a physical meaning and our aim is to obtain information about these parameters from
the observations. Inside the Bayesian framework for inference, our approach is the one
of parameter estimation. In the case at hand, the vector of parameters contains the six
free parameters of the CLUMPY model already described in \S\ref{sec:clumpy_models}.

In general, due to the presence of noise in the observations, any inversion 
procedure is not complete by just giving the values of the model parameters that better 
fit the observations. The full inference problem has to provide the posterior probability distribution 
function (PDF) $p(\thetabold|D,I)$ that describe the probability that a given set of parameters $\thetabold$
is compatible with the observables $D$ given the set of background a-priori assumptions $I$. As a consequence, 
statistically relevant information can be obtained from this PDF by marginalization 
(integration) of unimportant parameters. Specifically, the marginalization of all the parameters
except the one in which we are interested will give the probability distribution function of the parameter taking
into account all possible compatible values of the rest of parameters. Therefore, ambiguities and
degeneracies in the parameters are translated into marginal posterior distributions with
heavy tails. The cornerstone of the Bayesian approach to inference is the Bayes theorem, that
relates the posterior distribution $p(\thetabold|D,I)$ with any a-priori knowledge and the
information introduced by the data:
\begin{equation}
p(\thetabold|D,I) = \frac{p(D|\thetabold,I) p(\thetabold|I)}{p(D|I)},
\label{eq:bayes_theorem}
\end{equation}
where $p(D|I)$ is the so-called \emph{evidence}, $p(\thetabold|I)$ is denominated the 
\emph{prior} distribution and $p(D|\thetabold,I)$ is the so-called \emph{likelihood}.

The evidence, equal to the integral of
the posterior distribution over the parameter space, plays no role in the context of parameter
estimation because it is a constant that does not depend on the
model parameters $\thetabold$. However, it turns out to be crucial in the context of 
model selection. Strictly speaking, when one carries out parameter estimation in
the Bayesian context, one should indicate the estimation of the parameters, the error bars
and the value of the evidence. This way, other researchers can compare their results
with already published ones. The main difficulty is that the evidence is very difficult
to calculate and specifically designed algorithms are needed \citep{skilling04,gregory05}.

The prior distribution, $p(\thetabold|I)$, contains all relevant 
\emph{a-priori} information about the parameters of the model. Usually,
unless some information is available about the value of some parameters, it is common
to use uninformative priors like bounded uniform distributions or Jeffreys' priors
\citep[e.g.,][]{gregory05}. In case our a-priori knowledge of a parameter is 
sufficient to better constrain the value of a parameter, the Bayesian approach can
easily introduce the information into the inference process by appropriately
setting the prior distribution $p(\thetabold|I)$. Among the options, one can 
select exponential distributions if large values of the parameters are less
probable, Gaussian distributions around a given value with
a certain width if there is a region of the space of parameters with
more probability, etc. 

Finally, the likelihood $p(D|\thetabold,I)$ is a distribution
that gives the probability that the observed data set has been obtained using the set
of parameters $\thetabold$. Assuming that the observables are represented 
by the vector $\mathbf{d}$ of length $N$ and that the
model $M$ evaluates to the vector $\mathbf{y}$ of the same length contaminated by a 
noise component $\mathbf{e}$, we can write:
\begin{equation}
y_i = d_i + e_i, \qquad \forall i.
\end{equation}
When the chosen model parameters exactly correspond to those of the observed data set, the 
distribution of differences $y_i-d_i$ has to follow the distribution of the noise. Assuming that 
the noise is Gaussian distributed with a variance described by the vector $\sigmabold^2$, 
the likelihood function is given by:
\begin{equation}
p(D|\thetabold,I) \propto \prod_{i=1}^N  \exp \left[ \frac{(y_i-d_i)^2}{\sigma_i^2} \right],
\label{eq:likelihood}
\end{equation}
although other distributions can be used. For instance, in cases where only the upper of 
lower limit is known with a given confidence, it is possible to use skewed likelihood 
functions that appropriately take this into account. For simplicity, we assume
one-sided Gaussian likelihoods centered at zero in which $\sigma_i$ is adjusted so that the ratio of the 
integral of the likelihood between 0 and $d_i$ and its total area equals the confidence level of the observation.

Summarizing, Eq. (\ref{eq:bayes_theorem}) states that the 
probability that a model $M$ becomes plausible after
the data $D$ has been taken into account (posterior) depends on how plausible the model was before
presenting the data (prior) and how well the model fits the data (likelihood).

\subsection{Technicalities: the Markov Chain Monte Carlo method}
In order to calculate the posterior probability distribution function for 
one parameter and give estimations and confidence intervals, we have
to marginalize (integrate out) the rest of parameters from the full posterior
distribution:
\begin{equation}
p(\theta_i|D,I) = \int \mathrm{d}\theta_1\mathrm{d}\theta_2 \cdots \mathrm{d}\theta_{i-1} \mathrm{d}\theta_{i+1} \cdots 
\mathrm{d}\theta_{N_\mathrm{par}} p(\thetabold|D,I).
\end{equation}
To this end, \B\ utilizes a Markov Chain 
Monte Carlo \citep[MCMC;][]{metropolis53,neal93,gregory05} scheme based on the Metropolis algorithm.
The output of the MCMC method is a chain of models whose probability distribution 
follows the posterior distribution function. The MCMC technique can be also considered as an
integration method that returns marginal probability distribution for each parameter in
the model. As a consequence, the converged final 
Markov Chain obtained for each parameter automatically gives, after making histograms, its 
marginal posterior distribution (integrating out the rest of model parameters). 
Technically, our MCMC method works by proposing models using a multivariate Gaussian proposal distribution 
and accepting or rejecting the proposed models based on a standard Metropolis
acceptance rule. The proposal density distribution used in the first steps of the
chain is a multivariate Gaussian with diagonal covariance matrix that is set to 10\% of the 
allowed range of variation of the parameters. After a configurable initial period, the proposal density is changed
to a multivariate Gaussian with a covariance matrix that is estimated from the previous 
steps of the chain. In order to improve convergence, the covariance matrix is modified 
by a parameter that assures that the acceptance rate of proposed models is close to 25\%, a 
value that is close to the theoretical optimal value for simple problems \citep{gelman96}.

Although MCMC methods represent a huge step forward in the practical application of
Bayesian methods to the inference, one of the most important drawbacks is the necessity to sample
the whole posterior probability distribution, something that can be very time consuming.
Typical MCMC runs require Markov Chains of the order of 10$^4$-10$^5$ steps in order to
end up with correctly converged chains. In case the
evaluation of the forward model is not negligible, the total time can be
quite large and the systematic analysis of different observations remains completely unpractical.

\section{Principal Component Analysis}
\label{sec:appendix_pca}
Let us define $\mathbf{O}$ as the $N_\mathrm{models}\times N_\lambda$ matrix containing the wavelength variation 
of all the theoretical SEDs of the database where the mean SED has been 
substracted. The principal components can be found as the 
eigenvectors of this matrix of observations, so that they can be obtained by just
diagonalizing the matrix $\mathbf{O}$. Since we have $N_\mathrm{models} \gg N_\lambda$, this matrix is not square by 
definition and the dimension of the matrix can be so large that
computational problems can arise. However, it can be demonstrated that the 
right singular vectors of the matrix $\mathbf{O}$ are equal to the singular 
vectors of the covariance matrix:
\begin{equation}
\mathbf{X}=\mathbf{O}^\dag \ \mathbf{O},
\label{eq:correlation}
\end{equation}
that can be calculating with the Singular Value Decomposition algorithm 
\cite[SVD; see, e.g.,][]{numerical_recipes86}. The matrix $\mathbf{X}$ is 
the $N_\lambda \times N_\lambda$ covariance matrix and the superindex
$\dag$ represents the transposition operator. The same applies to the left singular
vectors, which are also eigenvectors of the covariance matrix $\mathbf{X}'=\mathbf{O} \ \mathbf{O}^\dag$.
The matrix $\mathbf{X}'$ has dimensions $N_\mathrm{models} \times N_\mathrm{models}$ and is typically much larger than 
the matrix $\mathbf{X}$. However, both descriptions are 
completely equivalent. The $i$-th principal components, $\vec{b}_i$, fulfills:
\begin{equation}
\mathbf{X} \vec{b}_i = k_i \vec{b}_i,
\end{equation}
with $k_i$ its associated eigenvalue. Writing all the eigenvectors as column vectors in the
unitary matrix $\mathbf{B}$ of dimensions $N_\lambda \times N_\lambda$, since this
represents a complete basis of the database, the observations can be
written as a linear combination of them as follows:
\begin{equation}
\mathbf{O}=\mathbf{C} \ \mathbf{B}^\dag,
\label{ec2}
\end{equation}
being $\mathbf{C}$ the $N_\mathrm{models}\times N_\lambda$ matrix of coefficients, whose $C_{ij}$ element
represents the projection of the observation $i$ on the eigenvector $j$:
\begin{equation}
\mathbf{C}=\mathbf{O} \ \mathbf{B}.
\label{ec3}
\end{equation}

The dimensionality reduction is carried out by using the matrix $\mathbf{\hat{B}}$ that 
contains only the first $N_\mathrm{red}$ eigenvectors that have been retained as 
containing the majority of signatures, so that, we end up with the
matrix of approximate SEDs:
\begin{equation}
\mathbf{\hat{O}} = \left( \mathbf{O} \ \mathbf{\hat{B}} \right) \mathbf{\hat{B}}^\dag.
\label{eq:reconstruction_approx}
\end{equation}

\section{Artificial Neural Network}
\label{sec:appendix_ann}
The function that the ANN whose topology corresponds to that shown in 
Fig. \ref{fig:neural_network} proposes depends on the six-dimensional vector of 
parameters $\thetabold=(\sigma, Y, N, q, \tau_V, i)$ and can be represented as:
\begin{equation}
C_{km} = f_m(\thetabold_k),
\label{eq:ann_functional}
\end{equation}
where $f_m(\thetabold_k)$ are highly non-linear functions whose functional form is
given explicitly below in Eq. (\ref{eq:ann_function}). The subindex $k$ labels all CLUMPY models
while the subindex $m \leq N'$ labels all the artificial neural networks
built for approximating the projection of the SEDs on each PCA eigenvector.
Using Eqs. (\ref{eq:ann_functional}) and (\ref{eq:reconstruction_approx}), the likelihood
function given by Eq. (\ref{eq:likelihood}) can be written as:
\begin{equation}
p(D|\thetabold,I) \propto \prod_{i=1}^N  \exp \left[ \frac{(\sum_j f_j(\thetabold) b_{j}^i-d_i)^2}{\sigma_i^2} \right],
\end{equation}
where $b_j^i$ is the $i$-th wavelength points of the $j$-th PCA eigenvector.

The output of each neural network can be expanded to read:
\begin{equation}
\label{eq:ann_function}
C_{km} = \sum_{j=1}^{N_h} v_j(m) \sigma \left[ \sum_{l=1}^6 w_j^l(m) \theta_k^l + u_j(m)\right],
\label{eq:neural_network}
\end{equation}
where the weights $v_j(m)$, $w_j^l(m)$ and the bias $u_j(m)$ represent free parameters that are optimized
during the training process. The symbol $N_h$ stands for the number of neurons in the hidden layer
and this essentially determines the number of weights and biases or, in other
words, the complexity of the network.

The training of the ANN is done by modifying, for fixed $m$, the weights and biases until minimizing the 
the quadratic difference between the true value of $C_{km}$ and the values returned by the 
artificial neural network for all the models included in the training:
\begin{equation}
E_m = \sum_{l=1}^{N_\mathrm{train}} \left[ C_{lm} - C^\mathrm{ANN}_{lm} \right]^2,
\end{equation}
where $N_\mathrm{train}$ is the number of models included in the training. It is important
to point out that neural networks suffer from the well-known
problem of over-fitting if the number of weights is too large. When a neural network is over-fitting the
training set, it looses generality and starts to ``mimick'' the fine details of each sample (similar to
what happens when a set of noisy points is fitted with a very high-degree polynomial). Although over-fitting
can be overcome with the aid of Bayesian techniques \citep{mackay92_1,mackay92_2}, we prefer to
use the less refined method of having a validation set as explained in the main text.


\begin{thebibliography}{52}
\expandafter\ifx\csname natexlab\endcsname\relax\def\natexlab#1{#1}\fi

\bibitem[{{Antonucci}(1993)}]{Antonucci93}
{Antonucci}, R. 1993, \araa, 31, 473

\bibitem[{{Aretxaga} {et~al.}(1999){Aretxaga}, {Joguet}, {Kunth}, {Melnick}, \&
  {Terlevich}}]{Aretxaga99}
{Aretxaga}, I., {Joguet}, B., {Kunth}, D., {Melnick}, J., \& {Terlevich}, R.~J.
  1999, \apjl, 519, L123

\bibitem[{{Asensio Ramos}(2006)}]{asensio_ramos06}
{Asensio Ramos}, A. 2006, ApJ, 646, 1445

\bibitem[{{Asensio Ramos} {et~al.}(2007{\natexlab{a}}){Asensio Ramos},
  {Mart\'{\i}nez Gonz\'alez}, \& {Rubi\~no
  Mart\'{\i}n}}]{asensio_martinez_rubino07}
{Asensio Ramos}, A., {Mart\'{\i}nez Gonz\'alez}, M.~J., \& {Rubi\~no
  Mart\'{\i}n}, J.~A. 2007{\natexlab{a}}, A\&A, 476, 959

\bibitem[{{Asensio Ramos} {et~al.}(2007{\natexlab{b}}){Asensio Ramos},
  {Socas-Navarro}, {L\'opez Ariste}, \& {Mart\'{\i}nez
  Gonz\'alez}}]{asensio_dimension07}
{Asensio Ramos}, A., {Socas-Navarro}, H., {L\'opez Ariste}, A., \&
  {Mart\'{\i}nez Gonz\'alez}, M.~J. 2007{\natexlab{b}}, ApJ, 660, 1690

\bibitem[{{Auld} {et~al.}(2008){Auld}, {Bridges}, \& {Hobson}}]{auld08}
{Auld}, T., {Bridges}, M., \& {Hobson}, M.~P. 2008, \mnras, 387, 1575

\bibitem[{{Ballantyne} {et~al.}(2006){Ballantyne}, {Shi}, {Rieke}, {Donley},
  {Papovich}, \& {Rigby}}]{Ballantyne06}
{Ballantyne}, D.~R., {Shi}, Y., {Rieke}, G.~H., {Donley}, J.~L., {Papovich},
  C., \& {Rigby}, J.~R. 2006, ApJ, 653, 1070

\bibitem[{{Brewer} {et~al.}(2007){Brewer}, {Bedding}, {Kjeldsen}, \&
  {Stello}}]{brewer_oscillations07}
{Brewer}, B.~J., {Bedding}, T.~R., {Kjeldsen}, H., \& {Stello}, D. 2007, ApJ,
  654, 551

\bibitem[{{Brewer} \& {Lewis}(2006)}]{brewer_lensing07}
{Brewer}, B.~J., \& {Lewis}, G.~F. 2006, ApJ, 637, 608

\bibitem[{{Brott} \& {Hauschildt}(2005)}]{phoenix_gaia05}
{Brott}, I., \& {Hauschildt}, P.~H. 2005, in ESA Special Publication, Vol. 576,
  The Three-Dimensional Universe with Gaia, ed. C.~{Turon}, K.~S. {O'Flaherty},
  \& M.~A.~C. {Perryman}, 565--+

\bibitem[{{Carroll} {et~al.}(2008){Carroll}, {Kopf}, \&
  {Strassmeier}}]{carroll_fast08}
{Carroll}, T.~A., {Kopf}, M., \& {Strassmeier}, K.~G. 2008, A\&A, 488, 781

\bibitem[{{Cornish} \& {Crowder}(2005)}]{cornish05}
{Cornish}, N.~J., \& {Crowder}, J. 2005, \prd, 72, 043005

\bibitem[{{Cybenko}(1988)}]{cybenko88}
{Cybenko}, G. 1988, Approximation by Superpositions of a Sigmoidal Function,
  Tech. rep.

\bibitem[{{Efstathiou} \& {Rowan-Robinson}(1995)}]{Efstathiou95}
{Efstathiou}, A., \& {Rowan-Robinson}, M. 1995, \mnras, 273, 649

\bibitem[{{Elitzur} \& {Shlosman}(2006)}]{Elitzur06}
{Elitzur}, M., \& {Shlosman}, I. 2006, \apjl, 648, L101

\bibitem[{{Fendt} \& {Wandelt}(2007)}]{fendt_pico07}
{Fendt}, W.~A., \& {Wandelt}, B.~D. 2007, ApJ, 654, 2

\bibitem[{{Fritz} {et~al.}(2006){Fritz}, {Franceschini}, \&
  {Hatziminaoglou}}]{Fritz06}
{Fritz}, J., {Franceschini}, A., \& {Hatziminaoglou}, E. 2006, \mnras, 366, 767

\bibitem[{{Gelman} {et~al.}(1996){Gelman}, {Roberts}, \& {Gilks}}]{gelman96}
{Gelman}, A., {Roberts}, G.~O., \& {Gilks}, W.~R. 1996, in Bayesian Statistics
  5, ed. J.~M. {Bernardo}, J.~{Berger}, A.~{Dawid}, \& A.~{Smith}, 599

\bibitem[{{Granato} \& {Danese}(1994)}]{Granato94}
{Granato}, G.~L., \& {Danese}, L. 1994, \mnras, 268, 235

\bibitem[{{Granato} {et~al.}(1997){Granato}, {Danese}, \&
  {Franceschini}}]{Granato97}
{Granato}, G.~L., {Danese}, L., \& {Franceschini}, A. 1997, ApJ, 486, 147

\bibitem[{{Gregory}(2005)}]{gregory05}
{Gregory}, P.~C. 2005, Bayesian Logical Data Analysis for the Physical Sciences
  (Cambridge: Cambridge University Press)

\bibitem[{{H{\"o}nig} {et~al.}(2006){H{\"o}nig}, {Beckert}, {Ohnaka}, \&
  {Weigelt}}]{Honig06}
{H{\"o}nig}, S.~F., {Beckert}, T., {Ohnaka}, K., \& {Weigelt}, G. 2006, A\&A,
  452, 459

\bibitem[{{Kashyap} \& {Drake}(1998)}]{kashyap98}
{Kashyap}, V., \& {Drake}, J.~J. 1998, ApJ, 503, 450

\bibitem[{{Lenzen} {et~al.}(1998){Lenzen}, {Hofmann}, {Bizenberger}, \&
  {Tusche}}]{Lenzen98}
{Lenzen}, R., {Hofmann}, R., {Bizenberger}, P., \& {Tusche}, A. 1998, in
  Presented at the Society of Photo-Optical Instrumentation Engineers (SPIE)
  Conference, Vol. 3354, Proc. SPIE Vol. 3354, p. 606-614, Infrared
  Astronomical Instrumentation, Albert M. Fowler; Ed., ed. A.~M. {Fowler}, 606

\bibitem[{{Lewis} \& {Bridle}(2002)}]{lewis02}
{Lewis}, A., \& {Bridle}, S. 2002, \prd, 66, 103511

\bibitem[{{Lo\`eve}(1955)}]{loeve55}
{Lo\`eve}, M.~M. 1955, Probability Theory (Princeton: Van Nostrand Company)

\bibitem[{{MacKay}(1992{\natexlab{a}})}]{mackay92_2}
{MacKay}, D. J.~C. 1992{\natexlab{a}}, Neural Comp., 4, 448

\bibitem[{{MacKay}(1992{\natexlab{b}})}]{mackay92_1}
---. 1992{\natexlab{b}}, Neural Comp., 4, 415

\bibitem[{{Marconi} {et~al.}(2000){Marconi}, {Schreier}, {Koekemoer},
  {Capetti}, {Axon}, {Macchetto}, \& {Caon}}]{Marconi00}
{Marconi}, A., {Schreier}, E.~J., {Koekemoer}, A., {Capetti}, A., {Axon}, D.,
  {Macchetto}, D., \& {Caon}, N. 2000, ApJ, 528, 276

\bibitem[{{Meisenheimer} {et~al.}(2007){Meisenheimer}, {Tristram}, {Jaffe},
  {Israel}, {Neumayer}, {Raban}, {R{\"o}ttgering}, {Cotton}, {Graser},
  {Henning}, {Leinert}, {Lopez}, {Perrin}, \& {Prieto}}]{Meisenheimer07}
{Meisenheimer}, K., {Tristram}, K.~R.~W., {Jaffe}, W., {Israel}, F.,
  {Neumayer}, N., {Raban}, D., {R{\"o}ttgering}, H., {Cotton}, W.~D., {Graser},
  U., {Henning}, T., {Leinert}, C., {Lopez}, B., {Perrin}, G., \& {Prieto}, A.
  2007, A\&A, 471, 453

\bibitem[{{Metropolis} {et~al.}(1953){Metropolis}, {Rosenbluth}, {Rosenbluth},
  {Teller}, \& {Teller}}]{metropolis53}
{Metropolis}, N., {Rosenbluth}, A.~W., {Rosenbluth}, M.~N., {Teller}, A.~H., \&
  {Teller}, E. 1953, J. Chem. Phys., 21, 1087

\bibitem[{{Neal}(1993)}]{neal93}
{Neal}, R.~M. 1993, Probabilistic Inference Using Markov Chain Monte Carlo
  Methods (Dept. of Statistics, University of Toronto: Technical Report No.
  0506)

\bibitem[{{Nenkova} {et~al.}(2002){Nenkova}, {Ivezi{\'c}}, \&
  {Elitzur}}]{Nenkova02}
{Nenkova}, M., {Ivezi{\'c}}, {\v Z}., \& {Elitzur}, M. 2002, \apjl, 570, L9

\bibitem[{{Nenkova} {et~al.}(2008{\natexlab{a}}){Nenkova}, {Sirocky},
  {Ivezi{\'c}}, \& {Elitzur}}]{Nenkova08a}
{Nenkova}, M., {Sirocky}, M.~M., {Ivezi{\'c}}, {\v Z}., \& {Elitzur}, M.
  2008{\natexlab{a}}, ApJ, 685, 147

\bibitem[{{Nenkova} {et~al.}(2008{\natexlab{b}}){Nenkova}, {Sirocky},
  {Nikutta}, {Ivezi{\'c}}, \& {Elitzur}}]{Nenkova08b}
{Nenkova}, M., {Sirocky}, M.~M., {Nikutta}, R., {Ivezi{\'c}}, {\v Z}., \&
  {Elitzur}, M. 2008{\natexlab{b}}, ApJ, 685, 160

\bibitem[{{Ossenkopf} {et~al.}(1992){Ossenkopf}, {Henning}, \&
  {Mathis}}]{Ossenkopf92}
{Ossenkopf}, V., {Henning}, T., \& {Mathis}, J.~S. 1992, A\&A, 261, 567

\bibitem[{{Pier} \& {Krolik}(1992)}]{Pier92}
{Pier}, E.~A., \& {Krolik}, J.~H. 1992, ApJ, 401, 99

\bibitem[{{Press} {et~al.}(1986){Press}, {Teukolsky}, {Vetterling}, \&
  {Flannery}}]{numerical_recipes86}
{Press}, W.~H., {Teukolsky}, S.~A., {Vetterling}, W.~T., \& {Flannery}, B.~P.
  1986, Numerical Recipes (Cambridge: Cambridge University Press)

\bibitem[{{Radomski} {et~al.}(2008){Radomski}, {Packham}, {Levenson},
  {Perlman}, {Leeuw}, {Matthews}, {Mason}, {De Buizer}, {Telesco}, \&
  {Orduna}}]{Radomski08}
{Radomski}, J.~T., {Packham}, C., {Levenson}, N.~A., {Perlman}, E., {Leeuw},
  L.~L., {Matthews}, H., {Mason}, R., {De Buizer}, J.~M., {Telesco}, C.~M., \&
  {Orduna}, M. 2008, ApJ, 681, 141

\bibitem[{{Rebolo} {et~al.}(2004){Rebolo}, {Battye}, {Carreira}, {Cleary},
  {Davies}, {Davis}, {Dickinson}, {Genova-Santos}, {Grainge}, {Guti{\'e}rrez},
  {Hafez}, {Hobson}, {Jones}, {Kneissl}, {Lancaster}, {Lasenby}, {Leahy},
  {Maisinger}, {Pooley}, {Rajguru}, {Rubi{\~n}o-Martin}, {Saunders}, {Savage},
  {Scaife}, {Scott}, {Slosar}, {Sosa Molina}, {Taylor}, {Titterington},
  {Waldram}, {Watson}, \& {Wilkinson}}]{rebolo04}
{Rebolo}, R., {Battye}, R.~A., {Carreira}, P., {Cleary}, K., {Davies}, R.~D.,
  {Davis}, R.~J., {Dickinson}, C., {Genova-Santos}, R., {Grainge}, K.,
  {Guti{\'e}rrez}, C.~M., {Hafez}, Y.~A., {Hobson}, M.~P., {Jones}, M.~E.,
  {Kneissl}, R., {Lancaster}, K., {Lasenby}, A., {Leahy}, J.~P., {Maisinger},
  K., {Pooley}, G.~G., {Rajguru}, N., {Rubi{\~n}o-Martin}, J.~A., {Saunders},
  R.~D.~E., {Savage}, R.~S., {Scaife}, A., {Scott}, P.~F., {Slosar}, A., {Sosa
  Molina}, P., {Taylor}, A.~C., {Titterington}, D., {Waldram}, E., {Watson},
  R.~A., \& {Wilkinson}, A. 2004, \mnras, 353, 747

\bibitem[{{Robitaille} {et~al.}(2007){Robitaille}, {Whitney}, {Indebetouw}, \&
  {Wood}}]{robitaille07}
{Robitaille}, T.~P., {Whitney}, B.~A., {Indebetouw}, R., \& {Wood}, K. 2007,
  ApJS, 169, 328

\bibitem[{{Rousset} {et~al.}(1998){Rousset}, {Lacombe}, {Puget}, {Hubin},
  {Gendron}, {Conan}, {Kern}, {Madec}, {Rabaud}, {Mouillet}, {Lagrange}, \&
  {Rigaut}}]{Rousset98}
{Rousset}, G., {Lacombe}, F., {Puget}, P., {Hubin}, N.~N., {Gendron}, E.,
  {Conan}, J.-M., {Kern}, P.~Y., {Madec}, P.-Y., {Rabaud}, D., {Mouillet}, D.,
  {Lagrange}, A.-M., \& {Rigaut}, F.~J. 1998, in Presented at the Society of
  Photo-Optical Instrumentation Engineers (SPIE) Conference, Vol. 3353, Proc.
  SPIE Vol. 3353, p. 508-516, Adaptive Optical System Technologies, Domenico
  Bonaccini; Robert K. Tyson; Eds., ed. D.~{Bonaccini} \& R.~K. {Tyson},
  508--516

\bibitem[{{Rubi{\~n}o-Martin} {et~al.}(2003){Rubi{\~n}o-Martin}, {Rebolo},
  {Carreira}, {Cleary}, {Davies}, {Davis}, {Dickinson}, {Grainge},
  {Guti{\'e}rrez}, {Hobson}, {Jones}, {Kneissl}, {Lasenby}, {Maisinger},
  {{\"O}dman}, {Pooley}, {Sosa Molina}, {Rusholme}, {Saunders}, {Savage},
  {Scott}, {Slosar}, {Taylor}, {Titterington}, {Waldram}, {Watson}, \&
  {Wilkinson}}]{rubino_martin03}
{Rubi{\~n}o-Martin}, J.~A., {Rebolo}, R., {Carreira}, P., {Cleary}, K.,
  {Davies}, R.~D., {Davis}, R.~J., {Dickinson}, C., {Grainge}, K.,
  {Guti{\'e}rrez}, C.~M., {Hobson}, M.~P., {Jones}, M.~E., {Kneissl}, R.,
  {Lasenby}, A., {Maisinger}, K., {{\"O}dman}, C., {Pooley}, G.~G., {Sosa
  Molina}, P.~J., {Rusholme}, B., {Saunders}, R.~D.~E., {Savage}, R., {Scott},
  P.~F., {Slosar}, A., {Taylor}, A.~C., {Titterington}, D., {Waldram}, E.,
  {Watson}, R.~A., \& {Wilkinson}, A. 2003, \mnras, 341, 1084

\bibitem[{{Schartmann} {et~al.}(2008){Schartmann}, {Meisenheimer}, {Camenzind},
  {Wolf}, {Tristram}, \& {Henning}}]{Schartmann08}
{Schartmann}, M., {Meisenheimer}, K., {Camenzind}, M., {Wolf}, S., {Tristram},
  K.~R.~W., \& {Henning}, T. 2008, A\&A, 482, 67

\bibitem[{{Schreier} {et~al.}(1998){Schreier}, {Marconi}, {Axon}, {Caon},
  {Macchetto}, {Capetti}, {Hough}, {Young}, \& {Packham}}]{Schreier98}
{Schreier}, E.~J., {Marconi}, A., {Axon}, D.~J., {Caon}, N., {Macchetto}, D.,
  {Capetti}, A., {Hough}, J.~H., {Young}, S., \& {Packham}, C. 1998, \apjl,
  499, L143

\bibitem[{{Skilling}(2004)}]{skilling04}
{Skilling}, J. 2004, in American Institute of Physics Conference Series, Vol.
  735, American Institute of Physics Conference Series, ed. R.~{Fischer},
  R.~{Preuss}, \& U.~V. {Toussaint}, 395

\bibitem[{{Skumanich} \& {L{\'o}pez Ariste}(2002)}]{skumanich02}
{Skumanich}, A., \& {L{\'o}pez Ariste}, A. 2002, ApJ, 570, 379

\bibitem[{{Telesco} {et~al.}(1998){Telesco}, {Pina}, {Hanna}, {Julian}, {Hon},
  \& {Kisko}}]{Telesco98}
{Telesco}, C.~M., {Pina}, R.~K., {Hanna}, K.~T., {Julian}, J.~A., {Hon}, D.~B.,
  \& {Kisko}, T.~M. 1998, in Presented at the Society of Photo-Optical
  Instrumentation Engineers (SPIE) Conference, Vol. 3354, Proc. SPIE Vol. 3354,
  p. 534-544, Infrared Astronomical Instrumentation, Albert M. Fowler; Ed., ed.
  A.~M. {Fowler}, 534--544

\bibitem[{{Trippe} {et~al.}(2008){Trippe}, {Crenshaw}, {Deo}, \&
  {Dietrich}}]{trippe08}
{Trippe}, M.~L., {Crenshaw}, D.~M., {Deo}, R., \& {Dietrich}, M. 2008, AJ, 135,
  2048

\bibitem[{{Tristram} {et~al.}(2007){Tristram}, {Meisenheimer}, {Jaffe},
  {Schartmann}, {Rix}, {Leinert}, {Morel}, {Wittkowski}, {R{\"o}ttgering},
  {Perrin}, {Lopez}, {Raban}, {Cotton}, {Graser}, {Paresce}, \&
  {Henning}}]{Tristram07}
{Tristram}, K.~R.~W., {Meisenheimer}, K., {Jaffe}, W., {Schartmann}, M., {Rix},
  H.-W., {Leinert}, C., {Morel}, S., {Wittkowski}, M., {R{\"o}ttgering}, H.,
  {Perrin}, G., {Lopez}, B., {Raban}, D., {Cotton}, W.~D., {Graser}, U.,
  {Paresce}, F., \& {Henning}, T. 2007, A\&A, 474, 837

\bibitem[{{Trotta}(2008)}]{trotta08}
{Trotta}, R. 2008, Contemporary Physics, 49, 71

\bibitem[{{Urry} \& {Padovani}(1995)}]{Urry95}
{Urry}, C.~M., \& {Padovani}, P. 1995, \pasp, 107, 803

\end{thebibliography}

\clearpage

\clearpage

\end{document}